\documentclass[final,5p,times,twocolumn,sort&compress]{elsarticle}
\pdfoutput=1
\usepackage{amssymb}
\journal{TBD}
\usepackage{hyperref}
\hypersetup{
  pdfauthor={John C. Travers},
  pdftitle={Optical Solitons in Hollow-Core Fibres},
  pdfpagemode=UseOutlines,
  colorlinks=true,
  hypertexnames=false,
  linkcolor=blue,
  urlcolor=blue,
  citecolor=blue}
\usepackage[all]{hypcap}
\usepackage{soul,color}
\usepackage{siunitx}
\DeclareSIUnit\bar{bar}

\newcommand{\etal}{et al.\@}
\newcommand{\eg}{\textit{e}.\textit{g}.\@}

\begin{document}
\begin{frontmatter}
\date{November 29, 2023}
\title{Optical Solitons in Hollow-Core Fibres}
\author{John C. Travers}
\ead{j.travers@hw.ac.uk}
\ead[url]{https://lupo-lab.com}
\affiliation{organization={School of Engineering and Physical Sciences, Heriot-Watt University},
            city={Edinburgh},
            postcode={EH14 4AS},
            country={United Kingdom}}

\begin{abstract}
I review the historical observation and subsequent research on optical soliton dynamics in gas-filled hollow-core optical fibres. I include both large-core hollow capillary fibres, and hollow-core photonic-crystal or microstructured fibres with smaller cores, in particular photonic bandgap and antiresonant guiding fibres. I discuss how the optical guidance properties of these different fibre structures influence the soliton dynamics that can be obtained. The dynamics I consider include: soliton propagation at peak power levels ranging from the megawatt to terawatt level, and pulse energies from sub-microjoule to millijoule range; pulse self-compression, leading to sub-cycle and sub-femtosecond pulse duration; soliton self-frequency shifting due to both the Raman effect, and the influence of photoionisation and plasma formation; and resonant dispersive wave emission, leading to the generation of tuneable few-femtosecond pulses across the vacuum and deep ultraviolet, visible, and near-infrared spectral regions.
\end{abstract}
\end{frontmatter}

\section{Introduction}
\noindent A soliton is a localised nonlinear wavepacket, or pulse, resulting from a balance between nonlinear and dispersive (or diffractive) effects in a system. They are shape-preserving, stable, propagate freely, and survive collisions between each other. These particle-like properties are the reason why Zabusky and Kruskal coined the particle-sounding term ``soliton''~\cite{zabuskyInteractionSolitonsCollisionless1965}. The first observation of soliton propagation was made by John Scott Russell almost 200 years ago~\cite{Russell1844}, in water waves travelling along the Union Canal near Edinburgh, very close to my research laboratories. Since then, soliton dynamics have been observed and studied in a wide range of physical systems~\cite{Scott1973,dauxois2010physics,Denschlag2000}. Soliton propagation in solid-core optical fibres was first proposed by Hasegawa and Tappert 50 years ago~\cite{Hasegawa1973,HASEGAWA2023129222}. By numerically solving the nonlinear Schr\"{o}dinger equation, they demonstrated that an anomalous group-velocity dispersion and a positive intensity-dependent nonlinear refractive index---achievable in an optical fibre---can give rise to stable temporal optical soliton propagation. Experimental confirmation was obtained in 1980 by Mollenauer \etal{}, who demonstrated not only non-dispersing soliton propagation, but also soliton self-compression \cite{Mollenauer1980}, an effect where the optical pulse duration decreases as the pulse propagates, without the need for any external phase compensation. This effect had been predicted in earlier theoretical work \cite{Zakharov1972,Satsuma1974}, and enables the generation of some of the shortest optical pulses ever produced, down to the sub-femtosecond timescale~\cite{Travers2019}. Subsequently, an extensive and rich variety of nonlinear optical dynamics involving solitons has been studied, as described elsewhere in many reviews and books~\cite{taylor_optical_2005,agrawal2019nonlinear,blanco2023bright,TAYLOR2023129382,GRELU2024130035,Tan2023}.

In this article, I review soliton dynamics in hollow-core optical fibres filled with gases. This platform has several properties that make it very interesting. First, gas-filled hollow-core fibres can be transparent in regions of the electromagnetic spectrum ranging from X-ray~\cite{watanabeSoftXrayTransmission1985} to the infrared \cite{daiHighpeakpowerPulsedCO21997} and beyond, extending far outside the transmission range of solid-core optical fibres. Second, many gases, particularly the light noble gases, have ionisation energies significantly higher than those of solid-state materials, allowing the transmission of much higher optical intensities before breakdown, and therefore higher optical pulse peak power and pulse energy can be transmitted and manipulated. Third, because gases are compressible, the material dispersion and nonlinearity of a gas-filled hollow fibre can be precisely tuned and adapted to achieve a specific set of dynamics, and also easily controlled during an experiment, something impossible in solid-core fibres. Fourth, the fundamental character of the nonlinear response can be altered through the type of gas used and the role of ionisation or Raman effects, enhancing the variety and richness of the nonlinear dynamics. These four properties have made nonlinear dynamics in gas-filled hollow fibres the focus of significant and enduring attention. Note that I do not cover optical solitons in liquid-filled hollow fibres, which have a different and fascinating behaviour~\cite{Chemnitz23,CHEMNITZ2023129874}.

\begin{figure*}[t!]
    \centering
    \includegraphics{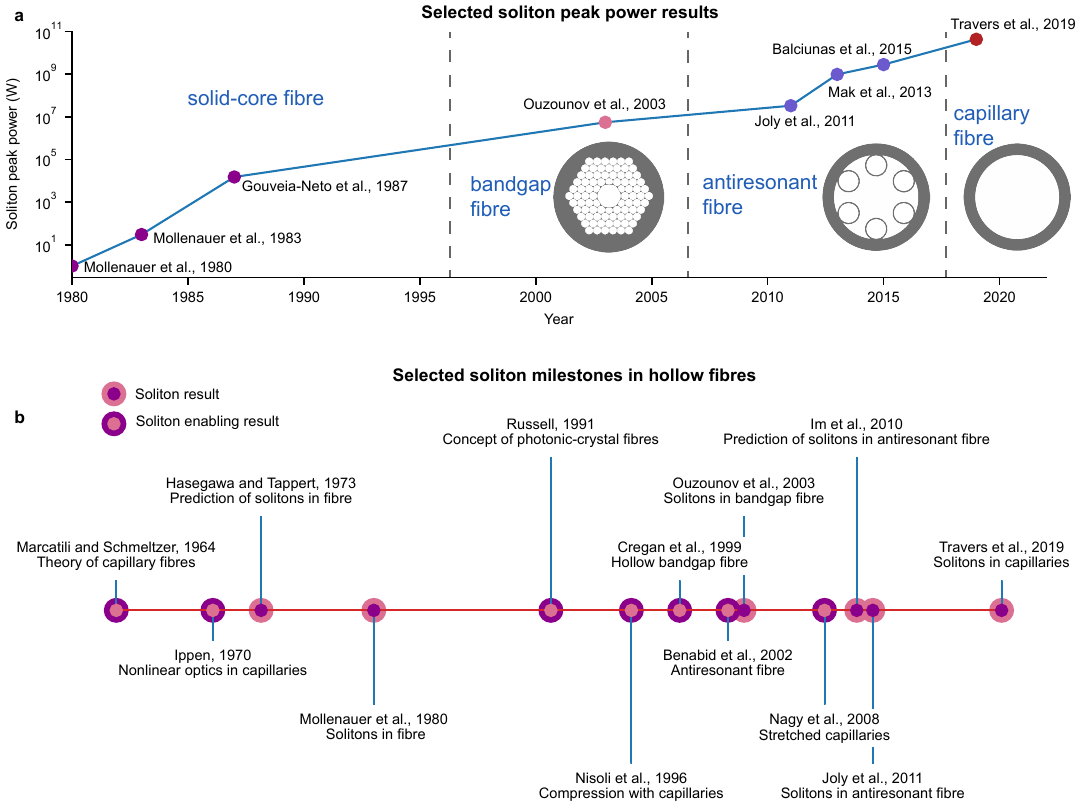}
    \caption{\label{fig:timeline} \textbf{Timeline of key developments in soliton propagation in hollow fibres.} (a) Evolution of soliton peak power achieved in fibres, along with illustrations of the main fibre structures (glass is grey, white represents hollow regions). (b) Milestones for solitons in hollow fibres along with important enabling fibre technologies. In both cases the selection of milestones is highly restricted due to space limitations---please see the main text for more comprehensive references. The references are: Marcatili and Schmeltzer, 1964~\cite{Marcatili1964}; Ippen, 1970~\cite{Ippen1970}; Hasegawa and Tappert, 1973~\cite{Hasegawa1973}; Mollenauer et al., 1980~\cite{mollenauer_experimental_1980}; Mollenauer et al., 1983~\cite{mollenauer_extreme_1983}; Gouveia-Neto et al., 1987~\cite{gouveia-netoGeneration33fsecPulses1987}; Russell, 2001~\cite{russellNeatIdeaPhotonic2001}; Nisoli et al., 1996~\cite{nisoli_generation_1996}; Cregan et al., 1999~\cite{cregan_single-mode_1999}; Benabid et al., 2002~\cite{benabid_stimulated_2002}; Ouzounov et al., 2003~\cite{ouzounov_generation_2003}; Nagy et al., 2008~\cite{nagyFlexibleHollowFiber2008}; Im et al., 2010~\cite{im_high-power_2010}; Joly et al., 2011~\cite{joly_bright_2011}; Mak et al., 2013~\cite{mak_two_2013}; Balciunas et al., 2015~\cite{balciunas_strong-field_2015}; Travers et al., 2019~\cite{Travers2019}.}
\end{figure*}

Optical solitons in hollow fibres are both intrinsically interesting---for example, they exhibit novel soliton self-frequency shifting regimes due to their interactions with photoionisation and plasma effects, or the molecular-structure dependent Raman response---and enable the creation of some unique light sources. They have enabled power scaling of optical solitons from the watt-level peak power first demonstrated in solid-core fibre, up to the sub-terawatt level (millijoule energy-scale) in hollow capillary fibres, as shown in Fig.~\ref{fig:timeline}(a). They have enabled optical pulse compression to sub-cycle and sub-femtosecond pulse duration, providing a simple and direct source of optical attosecond pulses. They have provided a route to generate few-femtosecond pulses tuneable across the vacuum and deep ultraviolet spectral region through resonant dispersive wave emission. And they facilitate ultra-broadband high-power supercontinuum formation.

The story of the observation and development of soliton propagation in hollow fibres is one of a competition between dispersion and attenuation. Gases have both a positive nonlinear refractive index and normal dispersion in the near-ultraviolet to mid-infrared spectral range. Therefore, the anomalous dispersion required to obtain bright temporal optical solitons must be provided by the hollow fibre itself.  The hunt for solitons in such fibres comes down to seeking a combination of waveguide structure, gas species, gas pressure, and pump pulse duration, to achieve sufficient anomalous dispersion while maintaining low loss propagation. Sufficient nonlinearity is usually easily achieved by increasing the pump power.

This article is structured as follows, with reference to the timeline illustrated in Fig.~\ref{fig:timeline}(b). First, I describe early nonlinear optics in the simplest hollow-core fibres: hollow-capillary fibres. I explain why observing solitons in such fibres was a challenge. Second, I review the pioneering work on soliton propagation in photonic bandgap fibres and the limitations of that type of waveguide. Third, I summarize the large amount of work on soliton propagation and related dynamics in antiresonant guiding hollow-core fibres---it was this platform that enabled soliton effects in gases to truly flourish. Finally, I revisit hollow-capillary fibres and describe the work that my group has pioneered on high-energy soliton self-compression and deep and vacuum ultraviolet generation. This was intended to be a concise review, aimed at providing a broad scope at a high level. I seem to have failed at the concise part, but I have indeed left out many details. To obtain these, I recommend reading the original literature cited, and also the many review papers on specific sub-topics~\cite{bhagwat_nonlinear_2008,benabid_linear_2011,travers_ultrafast_2011,russell_hollow-core_2014,Saleh2016,markosHybridPhotoniccrystalFiber2017}.

\section{\label{sec:presolhcf}Early non-soliton work in hollow capillary fibres}
Guiding electromagnetic waves inside hollow tubes was first analysed by Lord Rayleigh in 1887~\cite{struttPassageElectricWaves1887}, and practically utilised by Bose~\cite{emersonWorkJagadisChandra1997}, before being partially rediscovered multiple times~\cite{packardOriginWaveguidesCase1984a}, until a flurry of work in 1936~\cite{barrowTransmissionElectromagneticWaves1936,carsonHyperFrequencyWaveGuides1936,southworthHyperfrequencyWaveGuides1936}. This initial work focused on mm-waves and microwaves, but it is interesting to note that the idea of guiding light at optical wavelengths (\eg{} the ultraviolet, visible and near-infrared) in a hollow tube actually preceded it, considered for domestic lighting in 1881 by Wheeler~\cite{Wheeler1881}. Hollow metal ``pipes'' with a large core (\qty{25}{\mm}) were studied theoretically in 1962 for the purposes of optical communications by Eaglesfield~\cite{Eaglesfield1962}. However, the transmission and bend losses were too high to be practical. Hollow capillary fibres made from glass (see example structure in the right-most inset of Fig.~\ref{fig:timeline}(a)), along with small core hollow metallic waveguides, were subsequently studied in detail theoretically by Marcatili and Schmeltzer in 1964~\cite{Marcatili1964}.

Marcatili and Schmeltzer were interested in optical transmission for communications and in gas laser amplifiers. While the former turned out to be impractical in simple hollow capillaries, the latter is successful. Very soon after this, hollow capillaries started to be used in nonlinear optics experiments, for the same reason that optical fibres more generally are used: to maintain high intensity in a nonlinear medium over propagation lengths significantly exceeding those achievable in free space---the Rayleigh length of the focus. The earliest applications used liquid-filled capillaries: the first demonstration of Raman gain and lasing in an optical fibre~\cite{Ippen1970}, making use of a \qty{12}{\um} core diameter glass capillary filled with CS$_2$. In this case, the light is guided by total internal reflection because CS$_2$ has a higher refractive index than the glass used. Gas-based nonlinear optics followed, with enhancement of stimulated Raman scattering in 1976~\cite{Rabinowitz:76,milesCoherentStokesRaman1977}, and four-wave mixing in 1988~\cite{castillejoCoherentVacuumUltraviolet1988,heuerStimulatedRamanEffect1988}.

When filled with gas, hollow capillaries are not index guiding, because the refractive index of the gas is much lower than that of the glass cladding. While the analysis of Marcatili and Schmeltzer was a modal electromagnetic treatment, the very large core sizes (with respect to the wavelength) in hollow-core fibres means that good insight can be achieved with a light-ray picture, in which hollow capillaries guide by grazing incidence reflection. The angle of incidence between the light rays of the guided modes is so large that the reflection is exceedingly efficient. For example, for a typical \qty{250}{\um} core diameter hollow capillary, operating at \qty{800}{\nm}, the angle of incidence for a ray corresponding to the fundamental mode is $89.86^\circ$ and the power reflection coefficient is $0.99$. This ray picture can be used to derive exactly the same power attenuation coefficient that Marcatili and Schmeltzer obtained from a modal analysis~\cite{Marcatili1964,Fokoua2023}
\begin{equation}
    \label{eqn:alpha}
    \alpha(\lambda) = \left ( \frac{u_{nm}}{2\pi} \right )^2 \frac{\lambda^2}{a^3} \frac{n_\mathrm{g}^2+1}{(n_\mathrm{g}^2 - 1) ^{\frac{1}{2}}},
\end{equation}
where $a$ is the hollow core radius, $\lambda$ is the wavelength, $n_\mathrm{g}$ is the refractive index of the cladding glass, and $u_\mathrm{nm}$ is the $m^{\text{th}}$ zero of the Bessel function $J_\mathrm{n-1}$. Taking $n_\mathrm{g}\approx1.45$ for silica across the visible to near-infrared, we obtain an approximate loss length---over which the power is reduced by $1/\mathrm{e}$---of
\begin{equation}
    \label{eqn:hcfloss}
    L_\mathrm{loss}^\mathrm{HCF}=\frac{1}{\alpha} \approx \frac{14}{u_{nm}^2}\frac{a^3}{\lambda^2}\approx 2.4\frac{a^3}{\lambda^2},
\end{equation}
where the last expression is for the fundamental HE$_{11}$ mode with $m=n=1$. The $a^3/\lambda^2$ dependence on the loss length directly comes from the fact that smaller cores lead to smaller angles of incidence of the light rays, causing lower reflection per bounce, and this, in combination with the smaller core size, requires more bounces per unit length. Specifically, the angle between the fibre axis and the ray scales as $\propto \lambda / a$, and the number of bounces per unit length scales as $\propto \lambda / a^2$.

For sufficiently large cores, the loss length becomes reasonable at optical wavelengths, \eg{} for a \qty{250}{\um} core diameter the loss length is \qty{7}{\m} at \qty{800}{\nm}, much longer than the lengths usually used in nonlinear experiments. In contrast, the bend loss is very high. A simple analysis by Marcatili and Schmeltzer showed that the loss of a capillary with bend radius $R$ scales with the inverse dependence of Eq.~\ref{eqn:alpha}, $\alpha_\mathrm{bend}\propto a^3/\lambda^2R^2$. The situation was subsequently shown to be even worse due to mode mixing~\cite{miyagiWavePropagationAttenuation1984}. As a result, in practical use for nonlinear optics, simple hollow capillaries must be kept perfectly straight. To achieve this, rigid capillaries were originally used, with practical lengths limited to $\sim\qty{1}{\m}$ due to manufacturing tolerances\footnote{There are a few remarkable exceptions to this, such as the work of Suda \etal{} \cite{sudaGenerationSub10fs5mJoptical2005}.}. In 2008 this constraint was significantly relaxed through the invention of stretched hollow capillary fibres by Nagy \etal{}~\cite{nagyFlexibleHollowFiber2008}, which we will return to in Section~\ref{sec:hisol}.

A seminal contribution to ultrafast optics was the development of a technique for high energy pulse compression---based on nonlinear spectral broadening in a gas-filled hollow capillary---by Nisoli \etal{} in 1996~\cite{nisoli_generation_1996,nisoli_compression_1997}. I refer to this as nonlinear post-compression, or just post-compression, to disambiguate it from soliton-effect self-compression. This work enabled the production of few-cycle laser pulses with sufficient energy to readily drive strong-field physics, and became the work-horse for producing the driving pulses for attosecond science~\cite{sansoneHighenergyAttosecondLight2011a,frankInvitedReviewArticle2012}. Subsequently, it has enabled the scaling of few-cycle pulses to the multi-millijoule energy level~\cite{sudaGenerationSub10fs5mJoptical2005}, and peak powers approaching the terawatt-scale~\cite{nagyGenerationAboveTW5cycle2020}. A comprehensive review of this technology was provided in Ref.~\cite{NagyReview2021}.

Beyond pulse compression, gas-filled hollow capillaries started to be used for ultrafast nonlinear optics. Work included further development of frequency conversion through four-wave mixing in 1997~\cite{durfeeiiiUltrabroadbandPhasematchedOptical1997,durfeeiiiIntense8fsPulse1999a}, high-harmonic generation in 1999~\cite{durfeePhaseMatchingHighOrder1999a,tamakiPhasematchedHighorderharmonicGeneration1999}, cross-phase modulation in 1999~\cite{zheltikovSelfandCrossphaseModulation1999}, ultrafast Raman effects in 2000~\cite{wittmannNewRegimeFspulse2000}, and vacuum ultraviolet generation in 2001~\cite{misoguti_generation_2001}. Due to the ability to manipulate very high energy laser pulses in an environment with broadband guidance and low dispersion, development of this technology remains a very active research field in ultrafast nonlinear optics.

But why was soliton propagation not observed? To answer this queation, we must first re-cap some fundamental soliton parameters in the context of hollow-core fibres.

\begin{figure*}[t!]
    \centering
    \includegraphics{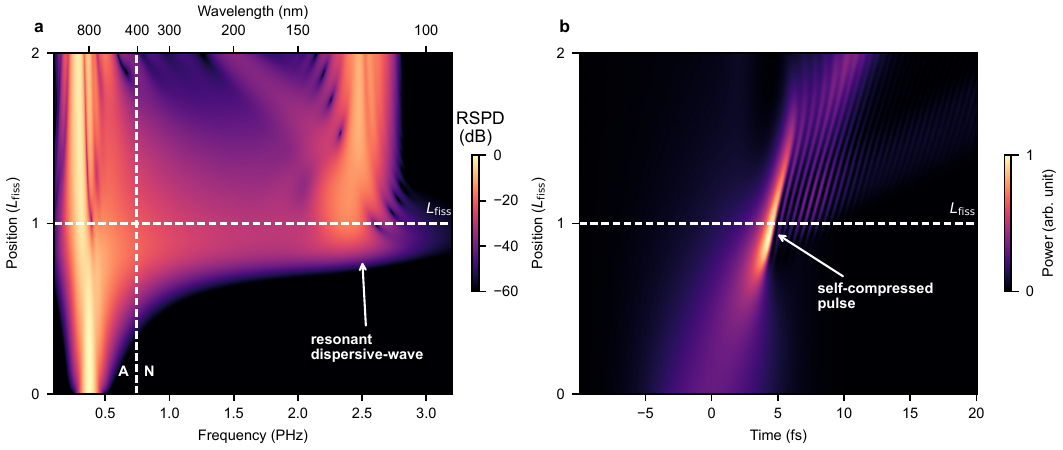}
    \caption{\label{fig:solevo} \textbf{Exemplary evolution of a self-compressing high-order soliton through a gas-filled hollow-core fibre.} For an \qty{8}{\fs}, \qty{800}{\nm} pulse, scaled to achieve $N=3$, and for a gas pressure such that $\lambda_\mathrm{zd}=\qty{400}{\nm}$. (a) Spectral power density. (b) Temporal power. The propagation is computed using my group's open-source code~\cite{brahms_lunajl_2021}. The horizontal dashed lines indicate the estimated soliton fission length $L_\mathrm{fiss}$. The vertical dashed line in (a) indicates the zero dispersion frequency, which separates the anomalous dispersion region (marked A) from the normal dispersion region (marked N). RSPD is relative spectral power density.}
\end{figure*}

The soliton order, $N$, is defined as
\begin{equation}
    \label{eqn:sol}
    N=\sqrt{\frac{L_\mathrm{disp}}{L_\mathrm{nl}}},
\end{equation}
where $L_\mathrm{disp} = \tau_0^2/|\beta_2|$, is the dispersion length, $\tau_0\approx\tau_\mathrm{fwhm}/1.763$ is the natural pump pulse duration, $\tau_\mathrm{fwhm}$ is the pulse full-width duration at half the peak power, $\beta_{2} = \delta^2\beta/\delta\omega^2$ is the group-velocity dispersion ($\beta$ is the axial propagation constant, and $\omega$ the angular frequency), $L_\mathrm{nl}=1/\gamma P_0$ is the nonlinear length, $P_0$ is the pump peak power, and $\gamma$ is the nonlinear coefficient~\cite{agrawal2019nonlinear}. To observe soliton effects, dispersion and nonlinearity must roughly balance, meaning that $L_\mathrm{disp}$ and $L_\mathrm{nl}$ must be similar, with $N=1$ corresponding to fundamental soliton propagation, where the phase induced from self-phase modulation is perfectly balanced by anomalous dispersion. Larger $N$ leads to soliton self-compression, where self-phase modulation initially dominates and broadens the pulse spectrum while inducing a chirp. The broader bandwidth increases the effect of anomalous dispersion, which compensates the chirp, producing a slight pulse compression. This shorter pulse---with higher peak power---then in turn induces yet more spectral broadening, leading to further pulse compression. This continuous cycle can lead to extreme self-compression, to a pulse duration even less than one optical cycle at the pump wavelength, and extreme spectral broadening, as depicted in Fig.~\ref{fig:solevo}. When $N$ is too large ($N \gtrsim 16$) modulational instability occurs, which drives an incoherent breakup of the pulse~\cite{TaiObservation1986,dudley_supercontinuum_2006}. While this is an exciting and important phenomenon in supercontinuum generation---including in hollow fibres~\cite{tani_phz-wide_2013}---and is closely related to soliton dynamics, I do not discuss it in this review, in the interests of conciseness.

The group-velocity dispersion of hollow capillary fibre is given by \cite{Marcatili1964,Travers2019}:
\begin{equation}
    \label{eqn:b2hcf}
    \beta_{2}(\lambda) \approx \frac{\lambda^3}{4 \pi c^2} \left( \rho \frac{\partial^2 \chi_\text{e}}{\partial \lambda^2} - \frac{u_{nm}^2}{2 \pi^2 a^2} \right),
    \label{eq:gvd}
\end{equation}
where $\rho$ is the pressure (and temperature) dependent gas density relative to some standard conditions, $\chi_\text{e}$ is the susceptibility of the filling gas species at those standard conditions, available through Sellmeier equations (such as \cite{Borzsonyi2008}), and $c$ is the speed of light in vacuum. The first term in Eq.~\ref{eqn:b2hcf} is due to the material dispersion of the filling gas, whereas the second term is the anomalous dispersion contribution due to confinement inside the waveguide core. One of the beautiful features of gas-filled hollow fibres of all kinds is that the dispersion landscape can be tuned simply by varying the filling gas pressure and species. This control has enabled many applications of these fibres to nonlinear optics~\cite{travers_ultrafast_2011,grigorovaDispersiontuningNonlinearOptical2023}.

For bright soliton propagation in a medium with positive nonlinear refractive index $n_2$, such as most gases in the visible to infrared range, the dispersion must be anomalous at the pump wavelength, or $\beta_{2}(\lambda_0) < 0$, where $\lambda_0$ is the pump wavelength. The sign of the dispersion switches at the zero dispersion wavelength, $\lambda_\mathrm{zd}$, defined by $\beta_2(\lambda_\mathrm{zd})=0$. In gas-filled hollow fibres, away from any structural resonances, the anomalous dispersion region extends from $\lambda_\mathrm{zd}$ to longer wavelengths.

The exact location of $\lambda_\mathrm{zd}$, combined with the pump wavelength, is a useful way to parameterize the dispersion landscape, even when changing other parameters, such as the core size, pump pulse duration, and even the fibre structure and gas species \cite{travers_ultrafast_2011,markosHybridPhotoniccrystalFiber2017,Travers2019,grigorovaDispersiontuningNonlinearOptical2023}. For example, it determines the wavelength at which resonant dispersive wave emission occurs (see Fig.~\ref{fig:solevo}), as well as the degree of soliton self-compression. If we fix $\lambda_\mathrm{zd}$ and analyse how the dispersion at other wavelengths varies with core size, we obtain $\beta_2(\lambda, \lambda_\mathrm{zd}, a) = \delta (\lambda, \lambda_\mathrm{zd})/a^2$, where the quantity $\delta(\lambda, \lambda_\mathrm{zd})$ is defined in Ref.~\cite{Travers2019} and only depends on the gas type and not on the gas pressure or any of the fibre parameters. From this expression, we can write the dispersion length as
\begin{equation}
    \label{eqn:ld}
    L_\mathrm{disp}^\mathrm{HCF} = \frac{\tau_0^2}{|\beta_2(\lambda)|} \propto a^2\tau_0^2.
\end{equation}
From Eq.~\ref{eqn:ld} we see explicitly that the importance of dispersion depends quadratically on the pump pulse duration and the core size.

Due to the large core size in conventional hollow capillaries (typically a diameter of around \qty{250}{\um} or more), high energy is required to achieve nonlinear effects, on the scale of at least several tens of microjoules, up to the multi-millijoule level. This is an advantage! The main reason to use such capillaries is for the energy and peak power scaling they enable compared to solid-core fibres. However, it does mean that they must be pumped with amplified laser systems, which is usually a Ti:sapphire\footnote{Optical parametric chirped-pulse amplifier systems (OPCPAs) can do better than this, but they were not widespread until recently, and hence not used to look for soliton effects in hollow capillaries.} laser system producing pulses of around \qty{30}{\fs} (unless using post-compression of the amplifier output, see Section~\ref{sec:hisol}). Therefore, the combination of large core size and relatively long pump pulse duration from high-energy laser amplifiers, means that dispersion is very weak, and the dynamics in hollow capillaries are dominated by nonlinearity, precluding soliton effects. Dispersion could be enhanced by shrinking the core size, but in that case, we face the growing importance of attenuation, as captured by Eq.~\ref{eqn:hcfloss}. As an example, the same \qty{250}{\um} core diameter HCF discussed above, which had a loss length of \qty{7}{\m} at \qty{800}{\nm}, has a dispersion length (when evacuated) of 46~m for a $35$~fs full-width at half-maximum (FWHM) duration pump pulse, suggesting that dispersion is negligible. Indeed, dispersion was justifiably neglected\footnote{Note that the dispersion experienced by pump pulses in hollow capillary post-compression experiments has been both normal and anomalous. The lack of observation of soliton effects was not due to the dispersion being normal, as is often claimed, but due to the fact that it was very weak, and for the pulse durations and fibre lengths used, it was negligible.} in post-compression experiments using hollow capillaries, and as a result, soliton dynamics were not observed.

To fully understand the conditions to observe soliton dynamics, we must also consider the nonlinearity and the length scale over which interesting soliton evolution occurs. For this purpose, we use the soliton fission length, as this is the length scale over which soliton self-compression takes place, leading to the shortest compressed pulse, the highest peak power, and maximum spectral broadening. This then initiates soliton shifting, resonant dispersive wave emission, and some forms of supercontinuum. Fig.~\ref{fig:solevo} illustrates how this process evolves during propagation, with distance normalized to the fission length. An approximate definition of this length\footnote{There are more precise definitions, for example \cite{dianov_optimal_1986,schadeScalingRulesHigh2021,cregoToolsNumericalModelling2023}, but understanding the general scaling with this simpler definition is most useful.} was provided by Dudley \etal{} in Ref.~\cite{dudley_supercontinuum_2006}:
\begin{equation}
    L_\mathrm{fiss} = \frac{L_\mathrm{disp}}{N}=\sqrt{L_\mathrm{disp}L_\mathrm{nl}}.
\end{equation}
The nonlinear coefficient inside a hollow capillary is determined purely by the pressure-dependent nonlinear refractive index of the filling gas---because the modal field has negligible intensity inside the glass structure~\cite{Marcatili1964}---and the effective area of the hollow mode, which, for the fundamental mode, is very well approximated by $A_\mathrm{eff}\approx 1.5a^2$~\cite{travers_ultrafast_2011}. It can therefore be written as
\begin{equation}
    \label{eqn:hcfgamma}
    \gamma = \frac{2\pi}{\lambda}\frac{\rho \bar{n}_2}{1.5a^2},
\end{equation}
where $\rho$ is the density of gas with respect to some standard conditions, and $\bar{n}_2$ is the nonlinear refractive index of the gas at those conditions. In Ref.~\cite{Travers2019} it was shown that if we fix $\lambda_\mathrm{zd}$ and $\lambda_0$, then the soliton fission length in a hollow core fibre scales as
\begin{equation}
    \label{eqn:hcffiss}
    L_\mathrm{fiss}^\mathrm{HCF}\propto \frac{\tau_\mathrm{fw} a^2}{\sqrt{I_0}},
\end{equation}
where $I_0$ is the peak intensity.

\begin{figure}[bt!]
    \centering
    \includegraphics{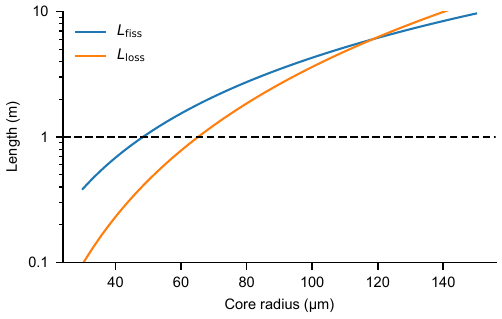}
    \caption{\label{fig:fisslong}\textbf{Soliton fission length in hollow capillary fibres with long pump pulses.} The soliton fission length $L_\mathrm{fiss}$ and loss length $L_\mathrm{loss}$ as a function of core radius for a hollow capillary filled with helium such that $\lambda_\mathrm{zd}=\qty{600}{\nm}$, pumped with \qty{35}{\fs} pulses at \qty{800}{\nm}. For each core size, the gas pressure is scaled to set $\lambda_\mathrm{zd}=\qty{600}{\nm}$, and the pump peak power is set to the maximum allowed intensity, following the limits determined in \cite{Travers2019}. For small core sizes $L_\mathrm{fiss} > L_\mathrm{loss}$, so attenuation dominates, and for larger core sizes $L_\mathrm{fiss} \gg \qty{1}{\m}$, precluding observation of soliton dynamics in conventional rigid capillaries.}
\end{figure}

Equations \ref{eqn:hcffiss} and \ref{eqn:hcfloss} are the basic scaling rules for soliton dynamics in hollow capillaries, and we can now answer our question: why was soliton propagation not observed?  To observe soliton dynamics in a fibre we must ensure that
\begin{equation}
    L_\mathrm{fiss} \lesssim L_\mathrm{loss},
\end{equation}
and also that $L_\mathrm{fiss}$ is shorter than practical capillary lengths.
As the peak intensity that can be propagated inside a hollow core fibre is limited by the breakdown of the filling gas due to ionisation~\cite{Travers2019}, and we need the gas to obtain nonlinearity and control the dispersion, our only means for controlling the fission length remain the pump pulse duration and the core size. To continue our example, taking parameters typical of a post-compression system~\cite{frankInvitedReviewArticle2012}, such as a \qty{250}{\um} core diameter filled with \qty{3}{\bar} of neon, \qty{800}{\nm}, \qty{35}{\fs} pump pulses with \qty{500}{\uJ} pulse energy, the calculated soliton fission length is \qty{4}{\m}, longer than the \qty{1}{\m} routinely employed, and hence soliton evolution was not observed. As shown in Fig.~\ref{fig:fisslong}, changing the core size does not help. For smaller core sizes, $L_\mathrm{fiss} > L_\mathrm{loss}$, so attenuation dominates, and for larger core sizes $L_\mathrm{fiss} \gg \qty{1}{\m}$, precluding observation of soliton dynamics in conventional rigid capillaries at this time. Much more recently, soliton dynamics in hollow capillaries have been achieved, but that story must wait until Section~\ref{sec:hisol}.

\begin{figure}[t!]
    \centering
    \includegraphics{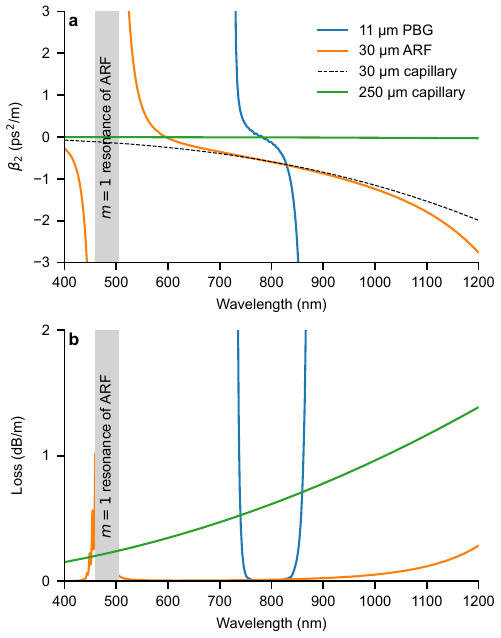}
    \caption{\label{fig:disp}\textbf{Linear properties of hollow-core fibres.} (a) Group velocity dispersion of three types of hollow-core fibre that have been used for soliton propagation: an \qty{11}{\um} core diameter photonic bandgap fibre (PBG) designed to operate around \qty{800}{\nm} (data from \cite{travers_ultrafast_2011}); a \qty{30}{\um} core diameter antiresonant fibre (ARF), with \qty{230}{\nm} wall thickness (calculated using \cite{rosaAnalyticalFormulasDispersion2021})---the black dashed curve shows a \qty{30}{\um} core diameter hollow capillary for comparison, and the grey region masks the resonance; and a \qty{250}{\um} core diameter hollow capillary fibre (calculated using \cite{Marcatili1964}). (b) The corresponding power loss curves for the same fibres (the \qty{30}{\um} core diameter hollow capillary is not shown in (b) as it is too lossy).}
\end{figure}

\section{Optical solitons in photonic bandgap fibres}
A route to achieve the required dispersion and loss balance to obtain soliton dynamics in a hollow fibre became available with the demonstration of single-mode hollow-core photonic bandgap fibres by Cregan \etal{} in 1999 \cite{cregan_single-mode_1999}, and substantially improved by Smith \etal{} in 2003 \cite{smithLowlossHollowcoreSilica2003}. The cladding structure in these fibres (see the inset figure in Fig.~\ref{fig:timeline}(a)) consists of a periodic arrangement of air-holes within glass, extending along the entire fibre in the direction of propagation, such that there is a bandgap for light incident on the periodic structure---out of the plane of periodicity---from inside a hollow core. Specifically, there are combinations of optical wavelength and angle of incidence which cannot propagate from air into the cladding, but which can propagate in a hollow core embedded within this structure, enabling the confinement of light~\cite{russellNeatIdeaPhotonic2001,Knight2003a,Russell2003,Russell2006}.

Photonic bandgap fibres are a type of photonic-crystal fibre, a concept developed by Philip Russell in 1991~\cite{russellNeatIdeaPhotonic2001}, and first demonstrated by him along with Tim Birks and Jonathan Knight in 1995, with the original aim to obtain photonic bandgap guidance. However, the technology developed, along with further insight, led to a generalisation of optical fibres with a wide variety of structures and wide-ranging guidance properties~\cite{Russell2006}.

Guidance in an air-core surrounded by a silica photonic-crystal cladding was first achieved in 1999 \cite{cregan_single-mode_1999}, and perfection of the fabrication process~\cite{smithLowlossHollowcoreSilica2003} delivered attenuation as low as 1.2~dB/km~\cite{Roberts2005}. Note that a distinction should be made between hollow-core photonic bandgap fibres, which have a full 2D out-of-plane bandgap, and Bragg~\cite{yehBraggReflectionWaveguides1976}, or Omniguide fibres~\cite{johnsonLowlossAsymptoticallySinglemode2001}, consisting of a 1D periodic annular array of rings of high and low refractive index, arranged concentrically around a central core. The latter has been mostly used in the infrared and THz spectral region, for example, for delivery of high-power CO$_2$ lasers. As far as I am aware no soliton dynamics have been demonstrated in that type of fibre.

\begin{figure*}[t!]
    \centering
    \includegraphics{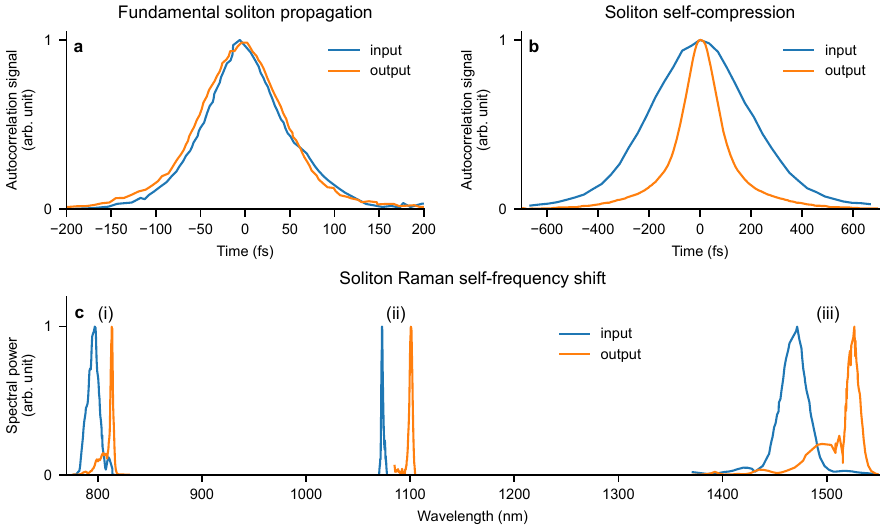}
    \caption{\label{fig:PBG} \textbf{Soliton results in hollow-core photonic bandgap fibres.} (a) Intensity autocorrelation traces of the input and output pulses from the first demonstration of soliton propagation in bandgap fibre, in this case a xenon-filled fibre pumped with \qty{75}{fs} pulses (data from \cite{ouzounov_generation_2003}). (b) Intensity autocorrelation traces of soliton self-compression of \qty{300}{\fs} pulses at \qty{540}{\nm} in a \qty{1}{\m} bandgap fibre (data from \cite{mosley_ultrashort_2010}). (c) Example spectra of Raman soliton self-frequency red-shift in bandgap fibre: (i) data from \cite{gerome_high_2008}; (ii) data from \cite{luan_femtosecond_2004}; (iii) data from \cite{ouzounov_generation_2003}.}
\end{figure*}

Photonic bandgap fibres introduced a new paradigm in optical fibre research. For the first time a small-core hollow fibre, which could be bent and coiled, offered low propagation losses and single-mode guidance, in strong contrast to hollow capillary fibres. Much immediate interest was in applications to telecommunications~\cite{polettiHighcapacityFibreopticCommunications2013a}, with the prospect of obtaining loss below the Rayleigh scattering limit of solid-core fibres, achieving low latency, and increasing bandwidth through lower dispersion and nonlinearity. This potential was never realized due to new loss mechanisms introduced by the photonic-crystal cladding structure, in particular surface-scattering losses~\cite{Fokoua2023}. Nevertheless, these fibres were quickly used for other applications: particle guidance~\cite{benabidParticleLevitationGuidance2002}, four-wave mixing~\cite{konorov_enhanced_2003}, stimulated Raman scattering in gases~\cite{benabid_ultrahigh_2004}, spectral broadening through self-phase modulation~\cite{konorov_self-phase_2004}, and infrared laser-pulse delivery~\cite{humbert_hollow_2004}.

Photonic bandgap fibres also solved the dispersion versus attenuation problem for soliton propagation, readily achieving $L_\mathrm{fiss} \ll L_\mathrm{loss}$. Bandgap fibres have three contributors to dispersion. The first two are similar to all hollow fibres, as expressed by Eq.~\ref{eqn:b2hcf}: the material dispersion of the gas filling the core, and the dispersion arising from confinement. However, it is the highly dispersive contribution from the optical properties of the bandgap cladding, which are dominant: at the edges of the bandgap the field penetrates further into the photonic-crystal cladding~\cite{robertsControlDispersionPhotonic2005}. An illustrative dispersion curve for a bandgap fibre is shown in Fig.~\ref{fig:disp} and shows a characteristic shape: the dispersion is strongly normal (positive) at the short wavelength band-edge and strongly anomalous (negative) at the long-wavelength band-edge, with a zero dispersion point within the band. The same figure shows the dispersion of a \qty{250}{\um} core diameter hollow capillary (a fairly typical core size)---which shows little dispersion---and an antiresonant guiding fibre (see Section~\ref{sec:arf}). While the antiresonant fibre (even one which has a larger core) is more dispersive across the central part of the band, the bandgap fibre has much stronger dispersion at the edges of the guidance window. In fact, one of the first applications of bandgap fibre was as a linear dispersive element for pulse compression (instead of prisms, gratings, or chirped mirrors) in all-fibre chirped-pulse amplification setups~\cite{de_matos_all-fiber_2003,de_matos_all-fiber_2004,limpert_all_2003}.

The nonlinear coefficient in a bandgap fibre is slightly enhanced compared to an equivalent-core hollow capillary, because the modal field extends into the glass structure with non-negligible intensity (due to the much higher nonlinear refractive index of the glass compared to the filling gas)~\cite{laegsgaardMaterialEffectsAirguiding2003a,hensleySilicaglassContributionEffective2007}. Nevertheless, it is still several orders of magnitude smaller than usual solid-core fibres, due to the larger core sizes and low nonlinearity of the filling gas. As an example, a typical bandgap fibre has $\gamma \sim\qty{5e-6}{\per\W\per\m}$ compared to $\gamma \sim\qty{5e-3}{\per\W\per\m}$ for solid-core fibre.

This greatly reduced nonlinearity, combined with the fact that the dispersion in a photonic bandgap fibre can be similar, or even exceed, that of a solid-core fibre, makes bandgap fibres, at first sight, an ideal platform for high peak-power soliton dynamics. However, bandgap fibres come with two severe restrictions: the guidance windows are very narrow, and the dispersion slope is very large; both are an inevitable consequence of the photonic bandgap mechanism. This restricts the spectral bandwidth, preventing the full range of soliton-related phenomena from being accessed, and consequently limits the shortest pulse durations that can propagate or be generated from soliton effect self-compression.

Soliton propagation in photonic bandgap fibre was first exploited by Ouzounov \etal{} in 2003 \cite{ouzounov_generation_2003}, who used a fibre filled with either air or xenon, and pumped at a wavelength of \qty{1510}{\nm}. When pumping the xenon-filled fibre with \qty{5.5}{\MW}, \qty{75}{\fs} pulses, they observed that the output pulses were almost unchanged despite propagating through \qty{1.7}{\m} of fibre, as shown in Fig.~\ref{fig:PBG}(a). This result was remarkable for several reasons. In addition to being the first demonstration of soliton dynamics in a hollow-core fibre of any kind, it was also the first ultrafast optics application of photonic bandgap fibres of any kind.

Higher-order soliton self-compression was also observed in bandgap fibres~\cite{konorov_self-compression_2005, ouzounov_soliton_2005}, but it was noted by Ouzounov \etal{} in 2005 that the large dispersion slopes characteristic of bandgap fibres limits high-ratio self-compression to input pulses with picosecond-scale duration~\cite{ouzounov_soliton_2005}. Self-compression was also observed in the green spectral region~\cite{mosley_ultrashort_2010}, shown in Fig.~\ref{fig:PBG}(b), with compression from $\sim\qty{300}{\fs}$ to $\sim\qty{100}{\fs}$ in a \qty{1}{\m} long fibre. Subsequently, the delivery and compression of \qty{1}{\ps}, \qty{5}{\uJ} pulses at \qty{1550}{\nm} and \qty{100}{\kHz} repetition rate (\qty{0.5}{\W}), was reported~\cite{peng_high_2011}.

When the fibre was filled with air instead of xenon, a Raman soliton self-frequency red-shift was observed, as shown in Fig.~\ref{fig:PBG}(c), similar to that observed in solid-core fibres \cite{dianov_stimulated-raman_1985,mitschke_discovery_1986,gordon_theory_1986}. In bandgap fibres, it is the molecular response of the nitrogen and oxygen in the air that provides the required Raman response~\cite{yan_impulsive_1985}, along with a contribution from the silica cladding \cite{luan_femtosecond_2004}. The work of Ouzounov \etal{} was swiftly followed by work from other groups, some more examples of which are also shown in Fig.~\ref{fig:PBG}(c) \cite{luan_femtosecond_2004,ivanov_frequency-shifted_2006,gerome_high_2008}. 

A frequency upshift is instead possible through the process of photoionisation and plasma formation. Fedotov \etal{} showed in 2007, both theoretically and experimentally, that ionisation in an air-filled fibre can cause a blue-shift of the input pulse \cite{fedotov_ionization-induced_2007}. A \qty{60}{fs} pump pulse at \qty{807}{\nm} with an energy of \qty{2.3}{\uJ} was upshifted more than \qty{10}{\THz}, which agreed with numerical simulations. The shift was described as a soliton self-frequency blue-shift. This work presaged the full development of the soliton self-frequency blue-shift in antiresonant guiding fibres (see Section~\ref{sec:ssfbs}).

Soliton self-compression occurs for $N > 1$. In contrast, \textit{adiabatic soliton compression} occurs when $N\approx 1$ but one or more of the dispersion, nonlinearity, or soliton energy is slowly adjusted (over the length scale of the soliton period $L_\mathrm{sol}=\pi L_\mathrm{disp}/2$) such that the soliton adapts to the new parameters by decreasing in temporal duration~\cite{kuehl_solitons_1988,chernikov_femtosecond_1991}. This approach was exploited in hollow-core bandgap fibre by Gérôme \etal{} in 2007 \cite{gerome_delivery_2007}, with an \qty{8}{m} tapered fibre. The taper reduces the hollow core diameter from \qty{7.2}{\um} to \qty{6.8}{\um}, which causes the dispersion at the \qty{800}{\nm} pump wavelength to vary from \qty{-0.027}{\ps^2\per\m} to close to \qty{0}{\ps^2\per\m}. At \qty{72}{\nano\joule} of pump energy, the \qty{195}{\fs} input pulses were compressed to \qty{90}{\fs} at the output. Subsequently, Welch \etal{} demonstrated adiabatic and soliton-effect self-compression using dispersion variation in a \qty{35}{\m} tapered bandgap fibre, with larger compression ratios, of \qty{2.5}{\ps} to \qty{215}{\fs} and \qty{1.2}{\ps} to \qty{175}{\fs}, at an output energy of \qty{5}{\nano\joule} and \qty{9.4}{\nano\joule} respectively~\cite{welch_solitons_2009}.

While these works pioneered the field of soliton dynamics in hollow-core fibre, the bandwidth restrictions inherent to bandgap guiding fibres severely restrict their use for the most extreme spectral broadening and pulse-compression effects.

\section{\label{sec:arf}Optical solitons in antiresonant fibres}

The bandwidth and dispersion-slope limitations inherent to photonic bandgap fibres were overcome through the use of a new generation of microstructured hollow fibres with a different guidance mechanism. Currently, the most common variations of this new kind of fibre are referred to by a variety of names, including: single-ring photonic-crystal fibre, negative-curvature fibre, revolver fibre, or antiresonant fibre. However, the first type of fibre within this class was the kagome fibre, so-called due to the fact that the cladding structure was similar to a kagome-lattice, originating from  a type of Japanese basket weaving. The first such fibre was demonstrated in 2002 by Benabid \etal{} \cite{benabid_stimulated_2002}. The fibre guided light over the entire visible to near-infrared spectral range, in stark contrast to photonic bandgap fibres. As the guidance mechanism of all of these types of hollow fibre is rooted in anti-resonance reflection---and their utility for harnessing soliton dynamics is almost identical---I refer to them all as antiresonant fibres. 

\subsection{Broadband guidance in antiresonant fibres}
There was considerable discussion in the literature about the possible guidance mechanisms of kagome fibre. It became clear that it could not be photonic bandgap guidance: while the cladding had a very low density of photonic states, it was not a true bandgap with zero photonic states~\cite{couny_generation_2007,Pearce2007,Argyros:07}. It was also found that the number of cladding layers was unimportant~\cite{Pearce2007, Argyros:07, Wang2011}, but that the thickness of the first air-glass interface surrounding the core, the core-wall, defined the guidance bands~\cite{Pearce2007,Argyros2008}, suggesting anti-resonance guidance~\cite{miyagiTransmissionCharacteristicsDielectric1980,Duguay1986,Litchinitser2002,Argyros2008}. In this way one can imagine the guidance mechanism as similar to hollow capillaries, through grazing incidence reflection from the core-wall, but enhanced in wavelength regions that are in anti-resonance with the thin core-wall~\cite{Litchinitser2002,Argyros2008}. The high-loss resonances are located at~\cite{Litchinitser2002,Pearce2007,yu_negative_2016}
\begin{equation}
    \label{eqn:arr}
    \lambda_m=\frac{2n_1d}{m}\sqrt{\left(\frac{n_2}{n_1}\right)^2-1}\approx\frac{2d}{m}\sqrt{n_2^2-1},
\end{equation}
where $n_1\approx 1$ is the low (core) refractive index, $n_2$ is the glass refractive index, $m$ is the order of the resonance, and $d$ is the thickness of the core-wall. Away from these resonant wavelengths, the confinement loss is strongly reduced. For example, a core-wall thickness of \qty{230}{\nm} would cause the first ($m=1$) resonance to occur at \qty{491}{\nm}, in agreement with Fig.~\ref{fig:disp}.

However, anti-resonance cannot be the full picture, as the experimentally and numerically modelled loss of these fibres was \textit{lower} than could be explained by anti-resonance reflection alone~\cite{Pearce2007,songQuantitativeAnalysisAntiresonance2019}. Furthermore, many cladding modes have propagation constants similar to the guided core modes. Answering the question of why this did not lead to high confinement loss was the idea that the very low modal overlap, due to highly dissimilar field profiles, between the core and cladding modes, inhibited the coupling between them~\cite{couny_generation_2007}. Similar to the anti-resonance picture, this cannot be the full story, as it does not capture the clear importance of the core-wall thickness. It is now widely accepted that both mechanisms are important~\cite{Fokoua2023}: anti-resonance enhanced grazing incidence reflection as the primary guidance mechanism, with inhibited coupling to the cladding modes acting to further reduce the confinement loss and reducing the bend loss.

The core wall of kagome fibres was found to benefit from curvature~\cite{Wang2011}. A greatly simplified hollow fibre design was then introduced by Pryamikov \etal{} in 2011~\cite{Pryamikov2011,Kolyadin2013}: a cladding formed by a single ring of smaller glass capillaries---which we call resonators. An example of this type of antiresonant fibre is shown in an inset figure in Fig.~\ref{fig:timeline}(a). This type of fibre was initially called single-ring or negative-curvature fibre, and subsequently named revolver fibre~\cite{Bufetov2018}, due to the fact that the cladding was a single-ring of circular tubes, reminiscent of the cylinder of a revolver. Negative-curvature fibres became a more generalised term, including the so-called ice-cream cone fibre \cite{Yu2012,yu_spectral_2013,yu_negative_2016}.  One explanation of the improved confinement loss for these was better spatial segregation of the core mode from the cladding structure, and in particular the nodes of the cladding. However, this does not explain the reduced loss in node-less hollow-core fibres. Instead, Murphy and Bird recently showed that these structures actually improve confinement loss through the concept of azimuthal confinement~\cite{murphyAzimuthalConfinementMissing2023}, and negative curvature is a by-product of structures with good azimuthal confinement.

Single-mode guidance can be optimised by carefully tuning the size of the cladding resonators to ensure high loss for higher-order modes, while keeping low confinement loss for the fundamental mode
\cite{Yu:16b,Uebel:16b,Newkirk:16,Hayes:17}.

Nested antiresonant fibres have additional resonators within the first ring, further reducing the confinement and bend loss~\cite{Belardi2014,Poletti2014}. Recently, nested antiresonant fibres have achieved attenuation values rivalling the best solid-core fibres~\cite{Fokoua2023} at telecommunications wavelengths, even though the fundamental guidance limits have not yet been reached. In an interesting twist from the original proposed light ``pipes'' of Eaglesfield~\cite{Eaglesfield1962}, nested antiresonant fibres are now the lowest loss optical waveguides ever made in all regions of the optical spectrum. In spectral regions that where glass is transparent, antiresonant hollow fibres---when optimised---enable lower transmission losses than solid-core fibres, because the hollow core suffers much less from Rayleigh scattering compared to glass~\cite{sakr2020hollow,Gao2020,osorio2023hollow,Fokoua2023}. Perhaps more significantly, antiresonant fibres are capable of guiding far outside the low-loss regions of the glass that forms the cladding; for example, deep into the mid-infrared~\cite{Pryamikov2011,Kolyadin2013,Yu2012,yu_spectral_2013, yu_negative_2016}, even beyond \qty{8000}{\nm}, or for vacuum ultraviolet light~\cite{belli_vacuum-ultraviolet_2015, ermolov_supercontinuum_2015}, even below \qty{120}{\nm}---although the fibre resonances in that region become very dense (as determined by Eq.~\ref{eqn:arr}).

\begin{figure}[t!]
    \centering
    \includegraphics{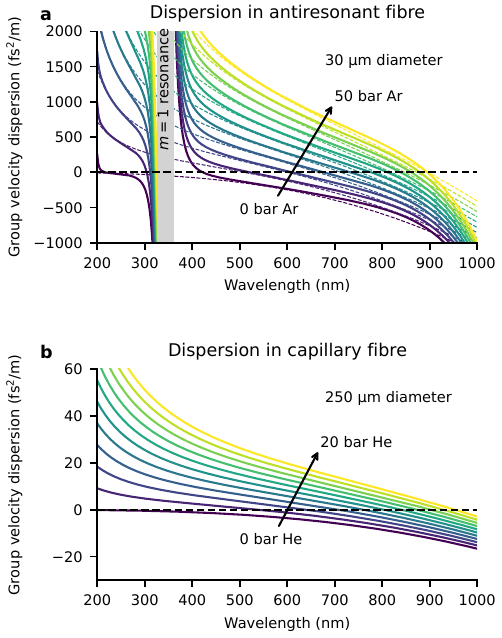}
    \caption{\label{fig:disptune}\textbf{Pressure tuning of dispersion in hollow-core fibres.} (a) Group velocity dispersion of a \qty{30}{\um} core diameter antiresonant fibre with \qty{160}{\nm} wall thickness filled with argon between \qty{0}{\bar} and \qty{50}{\bar} (calculated using \cite{rosaAnalyticalFormulasDispersion2021})---the thin dashed curves show the dispersion expected from a \qty{30}{\um} core diameter hollow capillary with the same gas filling, and the grey region masks the resonance. (b) Group velocity dispersion of a \qty{250}{\um} core diameter hollow capillary fibre filled with helium between \qty{0}{\bar} and \qty{20}{\bar} (calculated using \cite{Marcatili1964}).}
\end{figure}

\subsection{Dispersion and nonlinearity in antiresonant fibres}
The dispersion of antiresonant fibres is well described by the Marcatili and Schmeltzer model of hollow capillaries when operating far from the guidance resonances. This was first illustrated by Im \etal{} in 2009~\cite{im_guiding_2009}. One example of this is shown in Fig.~\ref{fig:disp}(a) (compare the orange and black dashed curves) and further illustrated in Fig.~\ref{fig:disptune}(a). But near the resonances the dispersion is strongly modified, leading to additional phase-matching for nonlinear processes~\cite{taniEffectAnticrossingsCladding2018,dengMicrojouleLevelMidInfraredFemtosecond,sollapurResonanceenhancedMultioctaveSupercontinuum2017a}. While truly accurate modelling of the dispersion still requires numerical modelling of Maxwell's equations (for example, using the finite element method), some semi-empirical models work well. In particular, the model of Rosa \etal{} \cite{rosaAnalyticalFormulasDispersion2021} provides both the confinement loss and dispersion, including resonance effects, and the model of Hasan \etal{} \cite{hasanEmpiricalFormulaeDispersion2018a} describes the dispersion and effective mode area of antiresonant fibres. The capillary approximation to dispersion in antiresonant fibres was first experimentally verified through phase-matched third harmonic generation~\cite{nold_pressure-controlled_2010}. As an example, the value of dispersion for a \qty{30}{\um} core diameter antiresonant fibre at \qty{800}{\nm} is $\sim\qty{-500}{\fs^2\per\m}$ and the dispersion length for a \qty{35}{\fs} pulse is \qty{67}{\cm}.

The nonlinear coefficient follows the same pressure-dependent scaling as hollow capillaries, Eq.~\ref{eqn:hcfgamma}, with negligible field strength in the glass---in contrast to bandgap fibres---and the effective area of the fundamental mode follows the same capillary approximation $A_\mathrm{eff}\approx 1.5a^2$. This was confirmed by two independent analyses \cite{rosaAnalyticalFormulasDispersion2021,hasanEmpiricalFormulaeDispersion2018a}. For a \qty{30}{\um} core diameter antiresonant fibre filled with \qty{10}{\bar} argon, the nonlinear coefficient is $\sim\qty{2e-6}{\per\W\per\m}$ at \qty{800}{\nm}.

Therefore, antiresonant fibres, especially with very thin core walls, $\sim\qty{200}{\nm}$ or less, such that the first resonance is shifted to the ultraviolet, well away from common pump laser wavelengths, have been used as essentially small-core capillaries with much lower loss~\cite{travers_ultrafast_2011}. The moderate anomalous dispersion in antiresonant fibres means that it can be almost ideally balanced with moderate gas-fill pressures, through Eq.~\ref{eqn:b2hcf}, leading to pressure-tunable dispersion across a wide wavelength range, as illustrated in Fig.~\ref{fig:disptune}(a); ideal for broadband ultrafast nonlinear optics at moderate energy (microjoule level), as established by Travers \etal{} in 2011~\cite{travers_ultrafast_2011}. Two non-soliton-related examples are the use of antiresonant fibres for efficient ultrafast four-wave mixing~\cite{belliHighlyEfficientDeep2019a}, and for the spectral broadening stage of conventional nonlinear post-compression (where the phase-compensation is achieved externally with chirped mirrors)~\cite{heckl_temporal_2011,emaury_beam_2013,mak_two_2013,Emaury:14,Mak2015,guichard_nonlinear_2015,gebhardt_nonlinear_2015,kottigEfficientSinglecyclePulse2020}. For more details of the many non-soliton applications of antiresonant fibres see Section~III in \cite{markosHybridPhotoniccrystalFiber2017}.

\begin{figure*}[t!]
    \centering
    \includegraphics{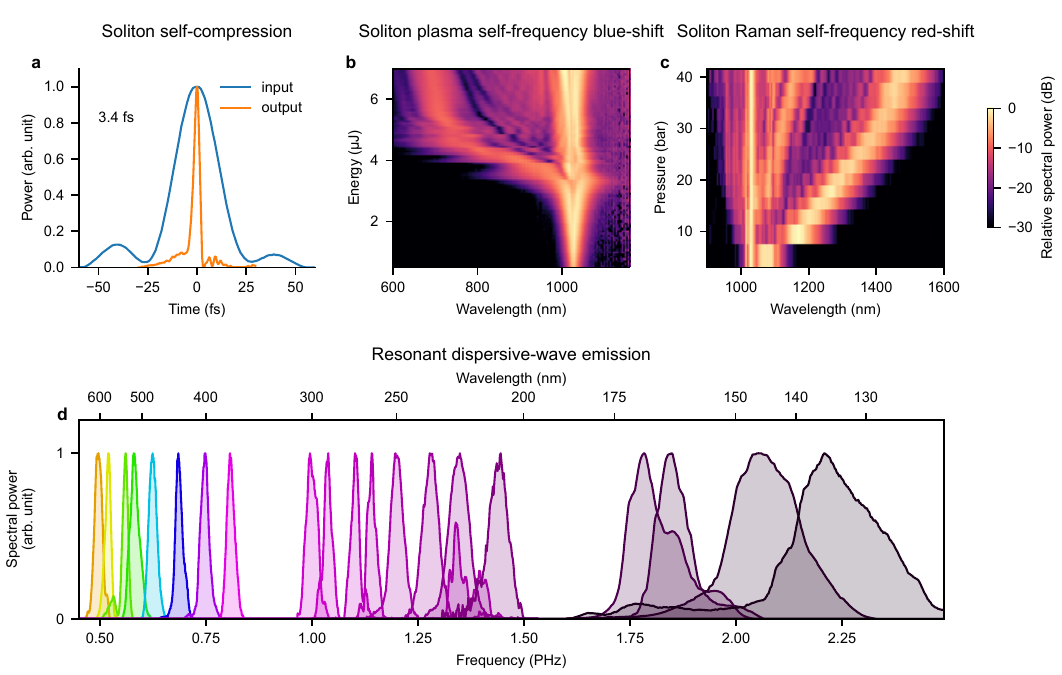}
    \caption{\label{fig:antiresonant} \textbf{Soliton results in antiresonant fibres.} (a) Soliton self-compression to \qty{3.4}{\fs} in a \qty{16.5}{\cm} long, \qty{59}{\um} core diameter antiresonant fibre filled with \qty{50}{\bar} neon and pumped with \qty{25}{\fs}, \qty{1030}{\nm} pulses (data from \cite{kottigEfficientSinglecyclePulse2020}). (b) Soliton plasma self-frequency blue-shift in a \qty{2}{\m} long, \qty{29}{\um} core diameter, argon-filled antiresonant fibre pumped with \qty{220}{\fs} pulses at \qty{1030}{\nm} (data from \cite{sabbahGenerationCharacterizationFrequency2023a}). (c) Soliton Raman self-frequency red-shift at \qty{8}{\MHz} repetition rate in a \qty{1.6}{\m} long, \qty{40}{\um} core diameter, hydrogen-filled fibre pumped with \qty{220}{\nano\J}, \qty{40}{\fs} pulses at \qty{1030}{\nm} (data from \cite{Tani2022}). (d) Tuneable resonant dispersive wave emission across the vacuum and deep ultraviolet using $\sim\qty{30}{\fs}$, \qty{800}{\nm} pump pulses and filling an antiresonant fibre with xenon, argon and helium (data from \cite{mak_tunable_2013,ermolov_supercontinuum_2015}).}
\end{figure*} 

\subsection{Soliton self-compression in antiresonant fibres}
The use of antiresonant fibres for enhanced soliton dynamics was proposed in a series of pioneering papers by Im \etal{} starting in 2009 \cite{im_guiding_2009,im_high-power_2010,im_soliton_2010}. Soliton self-compression and vacuum ultraviolet generation from resonant dispersive wave emission were both predicted. Experimental confirmation followed soon after by Joly \etal{} in 2011~\cite{joly_bright_2011}. They observed soliton self-compression of $\sim\qty{1}{\uJ}$, \qty{30}{\fs} pulses to \qty{9}{\fs} at \qty{800}{\nm}, with the lower duration limit due to their pulse-characterisation setup---numerical modelling and more recent experiments show that the compressed pulses were certainly shorter. In a wide-ranging review paper that analysed and predicted many possibilities for antiresonant fibres for nonlinear optics, Travers \etal{} predicted soliton self-compression to sub-femtosecond and sub-cycle pulses in 2011~\cite{travers_ultrafast_2011}.

As such short pulses are very challenging to measure, most demonstrations have relied on inference from other data, combined with numerical simulations. Sub-\qty{2}{\fs} pulses were inferred by Hölzer \etal{} in 2011~\cite{holzer_femtosecond_2011}, and Ermolov \etal{} in 2015~\cite{ermolov_supercontinuum_2015}, with estimated peak powers close to \qty{2}{\GW}. True confirmation of these results is awaiting sufficiently sophisticated pulse characterization measurements, although measurements of similarly extreme self-compression in hollow capillary fibres (see Section~\ref{sec:hisol}) lend credibility~\cite{Travers2019}. Actual measured results include those by Mak \etal{} in 2013~\cite{mak_two_2013}, who measured self-compression from \qty{24}{\fs} down to \qty{6.8}{\fs} at \qty{800}{\nm}, and Balciunas \etal{} in 2015~\cite{balciunas_strong-field_2015}, who compressed carrier-envelope phase stable \qty{35}{\uJ}, \qty{1800}{\nm} pulses to $\sim\qty{2.8}{\GW}$ peak power and \qty{4.5}{\fs} duration, which, accounting for the frequency upshift, is in the single-cycle regime. The latter pulses were rigorously characterised, proving experimentally for the first time that single-cycle pulses are achievable with self-compression in hollow-core fibres. In 2017, Gebhardt \etal{} demonstrated soliton self-compression of a high repetition-rate thulium-doped fibre laser system~\cite{gebhardtNonlinearPulseCompression2017}. They passed the output of their chirped-pulse amplification system---which produced \qty{110}{\fs}, \qty{41}{\uJ}, \qty{1900}{\nm} pulses at \qty{1.25}{\MHz} (corresponding to \qty{51}{\W} of average power)---through a \qty{42}{\cm} long, \qty{53}{\um} core diameter antiresonant fibre filled with argon with a pressure gradient from \qty{0.4}{\bar} to \qty{3}{\bar}. At the output they measured \qty{34.4}{\uJ} pulses with $\sim\qty{13}{\fs}$ duration, corresponding to \qty{1.4}{\GW}. In 2019, Ermolov \etal{} compressed \qty{1}{\uJ}, \qty{9}{\fs} pulses, from a carrier-envelope phase stable optical parametric chirped-pulse amplifier, to \qty{4.4}{\fs} and confirmed that the phase stability was preserved~\cite{ermolovCarrierenvelopephasestableSolitonbasedPulse2019a}. More recently, Savitsky\etal{} experimentally demonstrated the influence of the carrier-envelope phase of a multi-cycle pump pulse on the soliton self-compression dynamics and their interaction with third-harmonic generation~\cite{Savitsky:23}.

Soliton self-compression in the mid-infrared was demonstrated by Elu \etal{} in 2017~\cite{eluHighAveragePower2017}. They reported self-compression of \qty{97}{\fs} pulses to \qty{14.5}{\fs} at \qty{3250}{\nm}---which corresponds to 1.35 optical cycles---at a repetition rate of \qty{160}{\kHz} and corresponding average power of \qty{9.6}{\W}. In a further extension, Elu \etal{} in 2021~\cite{eluSevenoctaveHighbrightnessCarrierenvelopephasestable2021} compressed pulses with a duration of \qty{94}{\fs} and centre wavelength of \qty{3200}{\nm} to an estimated duration of \qty{3.5}{\fs}, forming a continuum spanning \qty{340}{\nm} to \qty{5000}{\nm}, using a \qty{0.2}{\m} long, \qty{92}{\um} core diameter antiresonant fibre filled with \qty{35}{\bar} argon.

Soliton self-compression has also been achieved when pumping at shorter wavelengths. In 2018, Hosseini \etal{}~\cite{hosseiniUVSolitonDynamics2018a}, used a \qty{0.7}{\m} long kagome fibre with a \qty{22}{\um} core diameter and core-wall thickness of just \qty{93}{\nm}. This thin core-wall pushes the first resonance down to around \qty{215}{\nm}, well away from the \qty{400}{\nm} pump wavelength. When filled with \qty{0.8}{\bar} argon and pumped with \qty{520}{\nano\joule}, \qty{40}{\fs} pulses at \qty{400}{\nm}, self-compression to below \qty{6}{\fs} was measured, based on back-propagation. This work was further extended, by Luan \etal{} in 2022~\cite{luanHighquality8foldSelfcompression2022}, who used even thinner core-walls of \qty{85}{\nm}, and added chirped mirrors after the fibre to ensure dispersion-compensated delivery of the self-compressed pulses from the fibre into the pulse characterisation setup. In this way, \qty{64}{\fs} pulses at \qty{400}{\nm} were measured to compress to \qty{7.6}{\fs} with \qty{2.4}{\uJ} of energy. 

To drive soliton self-compression with higher repetition-rate attention turned to ytterbium laser systems. In a pioneering work in 2015, Mak \etal{}~achieved soliton-effect self-compression of a ytterbium laser system to \qty{9.1}{\fs} at \qty{38}{\MHz}, with \qty{18}{\W} of average power. This is the highest repetition rate demonstrated to date. It was based on two-stage compression of a ytterbium thin-disk laser; the first stage was a conventional post-compression setup, and the second stage was used for self-compression. More recently, Köttig \etal{}~\cite{kottigEfficientSinglecyclePulse2020}, reported the construction of a similar system, but running at higher energy and lower repetition rate. They used a \qty{16.5}{\cm} long, \qty{59}{\um} core diameter antiresonant fibre filled with \qty{50}{\bar} neon and pumped with \qty{25}{\fs}, \qty{1030}{\nm} pulses to achieve near single cycle, \qty{3.4}{\fs}, \qty{5}{\uJ} pulses at \qty{9.6}{\MHz}; this result is shown in Fig.~\ref{fig:antiresonant}(a). In a detailed analysis in 2021, Schade \etal{}~\cite{schadeScalingRulesHigh2021}, studied analytically, numerically, and experimentally, the scaling of soliton self-compression dynamics in antiresonant fibres, with particular attention to inter-pulse plasma buildup effects. They identified an optimal parameter region, taking into account higher-order dispersion, photoionisation, self-focusing, and modulational instability.

Tani \etal{} recently demonstrated soliton self-compression to \qty{5}{\fs} of relatively low energy ($\sim\qty{200}{\nano\J}$) pulses at \qty{8}{\MHz} repetition rate~\cite{Tani2022}, and coupled this to driving soliton self-frequency red-shift due to Raman scattering, which I describe in more detail in Section~\ref{sec:ssfrs}. In 2023, Sabbah \etal{}~\cite{sabbahGenerationCharacterizationFrequency2023a} demonstrated self-compression of relatively long, \qty{220}{\fs}, pulses to \qty{13}{\fs} in a single and compact stage---consisting of a coiled \qty{2}{\m} long, \qty{29}{\um} core diameter, argon-filled antiresonant fibre---and then observed soliton self-frequency blue-shift due to photoionisation, which I describe in more detail in Section~\ref{sec:ssfbs}.

There are many applications of few-cycle and sub-cycle self-compressed pulses. One that was discussed at an early stage is for driving high-harmonic generation of extreme ultraviolet light or soft X-rays. A pioneering work in this direction was made by Heckl \etal{} in 2009~\cite{heckl_high_2009}; while this work did not harness soliton effects, it was the first demonstration of using hollow-core antiresonant fibre to enhance high-harmonic generation. Due to the tight confinement, the energy threshold was reduced to just \qty{200}{\nano\J} for the $7^\mathrm{th}$ harmonic at \qty{114}{\nm}. In 2011, Travers \etal{} proposed using soliton self-compression and phase-matching inside an antiresonant fibre to drive high-harmonic generation~\cite{travers_ultrafast_2011}, including moving to the short-wave infrared to extend the high-harmonic photon energy. The work of Fan \etal{} in 2014~\cite{fan_integrated_2014}, took the first steps in this direction, by reimaging the self-compressed pulse from an antiresonant fibre onto a gas jet to achieve high-harmonic generation up to \qty{50}{\eV}. In 2017, Tani \etal{} moved one step closer, by placing the gas jet directly at the output of an antiresonant fibre, to drive high-harmonic generation beyond \qty{45}{\eV} ($29^\mathrm{th}$ harmonic, at \qty{27.6}{\nm}), and made use of soliton dynamics to continuously tune the harmonic emission~\cite{taniContinuouslyWavelengthtunableHigh2017}. In 2021, Gebhardt \etal{} realised the full potential of this approach by directly generating high-harmonics inside the antiresonant fibre during soliton self-compression~\cite{gebhardtBrightHighrepetitionrateWater2021}. A careful construction of the laser system and tuning of the fibre design and gas-pressure gradient inside the fibre, to optimise the self-compression dynamics and phase-matching, led to rather impressive results: high-harmonic emission beyond \qty{300}{\eV} with exceptionally high flux due to the high repetition rate of \qty{98}{\kHz}.

In an interesting twist to the soliton self-compression dynamics, in 2017 Sollapur \etal{} \cite{sollapurResonanceenhancedMultioctaveSupercontinuum2017a} used propagation near one of the resonances to enhance the dispersion such that parameters that, under the simple capillary dispersion model, would have corresponded to a very high soliton order, and hence led to modulational instability, instead corresponded to a low enough soliton order for the pulse to undergo coherent soliton fission.

\subsection{Deep and vacuum ultraviolet generation in antiresonant fibres}
Solitons can emit a pulse of radiation at a phase-matched frequency through the process of resonant dispersive wave emission, which has been widely studied in solid-core fibre~\cite{wai_nonlinear_1986,karpman_radiation_1993,akhmediev_cherenkov_1995,skryabin_theory_2005,erkintalo_cascaded_2012,webb_generalized_2013,roger_high-energy_2013}. This process is illustrated in Fig.~\ref{fig:solevo}(a). The emission point is located at a frequency where the nonlinear propagation constant of the soliton matches the propagation constant of linear waves in the fibre,
\begin{equation}
    \beta(\omega)=\beta_\mathrm{sol}(\omega),
\end{equation}
where $\beta(\omega)$ is the linear propagation constant at angular frequency $\omega$, and $\beta_\mathrm{sol}(\omega)$ is the nonlinear propagation constant, at $\omega$, of the pump pulse centred at $\omega_\mathrm{sol}$,
\begin{equation}
    \beta_\mathrm{sol}(\omega) = \beta(\omega_\mathrm{sol}) + \beta_1(\omega_\mathrm{sol})[\omega - \omega_\mathrm{sol}] + \phi_\mathrm{nl},
\end{equation}
where $\phi_\mathrm{nl}$ is a nonlinear phase term, and $\beta_{n} = \delta^n\beta/\delta\omega^n$. Phase-matching is possible because solitons propagate without chirp---that is one of their defining characteristics, the balance between nonlinearity and dispersion---and so the frequency-dependent phase they acquire upon propagation differs from that of linear waves at the same frequency, but can match the phase of linear waves at a different frequency. If we use a Taylor expansion for the linear propagation constant around $\omega_\mathrm{sol}$, we find the phase matching condition becomes
\begin{equation}
    \sum_{n \ge 2}\frac{\beta_n(\omega_\mathrm{sol})}{n!}[\omega - \omega_\mathrm{sol}]^n = \phi_\mathrm{nl}.
\end{equation}
Therefore, the phase matching is mostly determined by high-order dispersion, with a nonlinear shift.

In hollow-core fibres, due to the relatively flat and smooth dispersion and broadband guidance, resonant dispersive wave emission can be tuned across the vacuum ultraviolet, deep ultraviolet and visible spectral region when pumping in the near-infrared. Efficient transfer of energy to the resonant dispersive wave requires spectral overlap of the driving pulse with the phase-matched point. For frequency shifts to the deep and vacuum ultraviolet from infrared driving pulses this requires a spectrum much larger than the pump pulse, and so resonant dispersive wave emission in hollow-core fibres usually occurs at the point of maximum self-compression of a high-order soliton, where the spectral broadening is maximum.

Resonant dispersive wave emission in antiresonant fibres was predicted by Im \etal{} in 2010 \cite{im_high-power_2010}, and experimentally demonstrated by Joly \etal{} in 2011~\cite{joly_bright_2011}. In the latter, an argon-filled, \qty{30}{\um} core diameter, kagome fibre was pumped with $\sim\qty{1}{\micro\joule}$, \qty{30}{\fs} pulses at \qty{800}{\nm}, and a deep ultraviolet resonant dispersive wave was observed to be pressure-tuneable from \qty{200}{\nm} to \qty{320}{\nm}, with a conversion efficiency up to 5\%. Joly \etal{} also identified the importance of self-steepening to enhance the efficiency of resonant dispersive wave emission. This work was extended by Mak \etal{} in 2013 \cite{mak_tunable_2013}, who tuned both the gas pressure and the gas species and achieved resonant dispersive wave emission between \qty{180}{\nm} and \qty{550}{\nm}, shown in Fig.~\ref{fig:antiresonant}(d). Extension further into the vacuum ultraviolet was first achieved by Belli \etal{} in 2015 \cite{belli_vacuum-ultraviolet_2015}, who generated a resonant dispersive wave at \qty{182}{\nm} which was then broadened down to \qty{124}{\nm} due to Raman-induced molecular modulation in the hydrogen filling gas (see Section~\ref{sec:ssfrs}). Direct tuning of resonant dispersive wave emission across the vacuum ultraviolet was subsequently achieved by Ermolov \etal{} also in 2015, \cite{ermolov_supercontinuum_2015}, who obtained emission down to \qty{113}{\nm} in helium, limited by the absorption of the MgF$_2$ windows used on the gas cells; some of those results are also shown in Fig.~\ref{fig:antiresonant}(d). It has been suggested with numerical simulations that further extension into the vacuum ultraviolet is feasible with tapered antiresonant fibres~\cite{habibMultistageGenerationExtreme2018,habibExtremeUVLight2019}, but this is yet to be experimentally demonstrated; only one experimental result in actively tapered antiresonant fibres has been demonstrated to date, of supercontinuum to the deep ultraviolet~\cite{sureshDeepUVenhancedSupercontinuumGenerated2021a}. Ermolov \etal{} also noted the importance of ionisation-driven soliton self-frequency blue-shift to further enhance the efficiency of resonant dispersive wave emission as predicted by Saleh \etal{} and discussed in Section~\ref{sec:ssfbs}. 

The use of gas pressure gradients between the input and output gas cells was first demonstrated by Mak \etal{} in 2013~\cite{mak_tunable_2013}. A decreasing pressure gradient enables the direct delivery of pulses into vacuum---studied in detail by Ermolov \etal{} \cite{ermolov_low_2013}---which is essential for preserving the shortest self-compressed or ultraviolet pulses. Pressure gradients also alter the nonlinear dynamics, by changing the nonlinearity, dispersion and hence the phase-matching condition along the fibre.

It is also possible for resonant dispersive waves to be emitted in higher-order modes, as demonstrated by Tani \etal{} in 2014~\cite{tani_multimode_2014}. In that case it is the linear propagation constant of the higher-order mode that must match the full nonlinear propagation constant of the self-compressing high-order soliton in the fundamental mode. This tends to lead to less efficient emission but to emission at a higher frequency.

The role of the fibre resonances on resonant dispersive wave emission was studied in detail by Tani \etal{} in 2018~\cite{taniEffectAnticrossingsCladding2018}. They showed that the large dispersion crossings induced by the fibre resonances give rise to additional phase-matched dispersive wave and four-wave mixing peaks. Furthermore, they showed that the fibre resonances can alter the emission of conventional ultraviolet resonant dispersive waves, and that their effect can be minimised by shifting the resonance wavelengths far from the pump.

Pumping gas-filled antiresonant fibres at longer wavelengths shifts the resonant dispersive wave tuning range. Utilizing the wavelength tuneability of an optical parametric amplifier, Cassataro \etal{} demonstrated resonant dispersive wave tuning around \qty{300}{\nm} through pump wavelength tuning (between \qty{1600}{\nm} and \qty{1800}{\nm}) for the first time (as opposed to gas pressure tuning)~\cite{Cassataro:17}. A similar result was reported by Adamu \etal{} in 2019 \cite{adamuDeepUVMidIRSupercontinuum2019}. In 2017, Meng \etal{} \cite{Meng:17}, drove resonant dispersive wave emission at around \qty{1000}{\nm} with a tunable pump wavelength in the range \qty{1300}{\nm} to \qty{1500}{\nm}. The much smaller frequency shift between the pump wavelength and the dispersive wave led to higher conversion efficiency than usual, up to 16\%. The self-compression and supercontinuum work of Elu \etal{} in 2021~\cite{eluSevenoctaveHighbrightnessCarrierenvelopephasestable2021} also exhibited resonant dispersive wave emission close to \qty{500}{\nm} when pumping in the mid-infrared at \qty{3200}{\nm}.

Moving to shorter pump wavelengths, in 2018 Hosseini \etal{}~\cite{hosseiniUVSolitonDynamics2018a} pumped resonant dispersive wave emission at \qty{400}{\nm} and achieved ultraviolet pulses tunable from \qty{195}{\nm} to \qty{260}{\nm}.

The repetition rate of ultraviolet dispersive wave emission was scaled by moving to ytterbium pump laser systems. In 2015, Mak \etal{}~\cite{Mak2015}, obtained dispersive wave emission tunable between \qty{380}{\nm} and \qty{520}{\nm} at \qty{38}{\MHz}---the highest repetition rate ever reported---with 4\% of the output, or \qty{500}{\mW} average power, in the resonant dispersive wave centred at \qty{420}{\nm}. In 2017, Köttig \etal{}~\cite{kottig2017generation}, used a \qty{53}{\um} core diameter fibre, filled with \qty{53}{\bar} helium, pumped with \qty{17}{\micro\joule} pulses at \qty{100}{\kHz}, to obtain \qty{1.05}{\micro\joule} energy pulses at \qty{205}{\nm}, corresponding to \qty{105}{\mW}, and 6.2\% conversion from the input pulses. Tuning the system to emit longer wavelengths allowed power-scaling: when using \qty{53}{\bar} neon and a repetition rate of \qty{1.92}{\MHz}, an average output power of \qty{1.03}{\W} was obtained at \qty{275}{\nm}. At lower repetition rates the efficiency was higher, with a total input-to-output efficiency of 8.2\% achieved at \qty{100}{\kHz}, with \qty{0.74}{\micro\joule} generated at \qty{264}{\nm}.

In a study aimed at broadening the spectrum of the ultraviolet dispersive wave, Smith \etal{}~\cite{Smith:20} turned to pump energy modulation. The core idea was that different pump pulse energies self-compress and generate resonant dispersive waves a slightly different wavelengths (due to the nonlinear contribution to the phase-matching). Therefore, by rapidly modulating the pump energy the spectrum around the average dispersive wave emission peak becomes smoother and broadened. Smith \etal{} demonstrated a dispersive wave bandwidth of \qty{52}{\nm} centred at \qty{324}{\nm}, which was more than double the unmodulated case.

One of the most attractive features of resonant dispersive wave emission is that the ultraviolet pulse duration is exceedingly short, with numerical predictions of just few-femtosecond duration. In 2015, Köttig \etal{}~\cite{kottig_high_2015} performed the first temporal characterization of deep ultraviolet resonant dispersive wave emission, putting an upper limit on the pulse duration of $<\qty{10}{\fs}$ at the fibre exit. However, this estimate was based on back-propagation through the air dispersion and gas-cell windows on route to the two-photon autocorrelator used for pulse duration measurements. As no phase information is available from the autocorrelator it was impossible to fully retrieve the ultraviolet pulse at the fibre exit. Subsequently, Ermolov \etal{} \cite{ermolov_multi-shot_2016}, used a transient-grating cross-correlation frequency-resolved optical gating system to completely characterise a \qty{4}{\fs} ultraviolet pulse duration at \qty{270}{\nm} with \qty{100}{\nano\J}; while this was also based on back-propagation, the full phase retrieval of the measurement made the back propagation more reliable. In 2019, Brahms \etal{}~\cite{brahmsDirectCharacterizationTuneable2019}, fully characterised ultraviolet resonant dispersive waves delivered directly in vacuum, without any back propagation, for the first time. They measured $\sim\qty{3}{\fs}$ duration from \qty{225}{\nm} up to \qty{300}{\nm}. Also in 2019, Ermolov \etal{} verified that carrier-envelope phase stability is transferred to resonant dispersive waves~\cite{ermolovCarrierenvelopephasestableSolitonbasedPulse2019a}, by beating the dispersive wave peak at \qty{415}{\nm} with the second-harmonic of the pump laser. In 2020, Adamu \etal{} \cite{adamuNoiseSpectralStability2020}, measured the noise properties of resonant dispersive wave emission, with results indicating very high values of noise. However, in that case a very noisy pump laser was used, far below the standard of state-of-the-art systems, and this seemed in contradiction to the phase stability demonstrated by Ermolov \etal{}. In a subsequent paper, Smith \etal{} \cite{smithLownoiseTunableDeepultraviolet2020}, used a lower noise ytterbium pump system, showing that low noise can be achieved, in good agreement with a later theoretical analysis \cite{brahmsTimingEnergyStability2021} and experimental measurements in hollow capillary fibres~\cite{Travers2019}.

This tunable source of few-femtosecond pulses has many applications, not least to ultrafast spectroscopy. In a first application, Bromberger \etal{} \cite{bromberger_angle-resolved_2015}, used tuneable vacuum ultraviolet pulses, down to \qty{140}{\nm}, for angle-resolved photoemission spectroscopy. In a second application, Kotsina \etal{} \cite{kotsinaUltrafastMolecularSpectroscopy2019} used resonant dispersive wave emission for ultrafast pump-probe spectroscopy in time-resolved photoelectron imaging experiments.

Recently, there has been interest in energy down-scaling deep ultraviolet generation by making use of small-core antiresonant fibres. In 2021, Xiong \etal{} \cite{xiongLowenergythresholdDeepultravioletGeneration2021}, demonstrated deep ultraviolet resonant dispersive wave generation in an antiresonant fibre with a core diameter of \qty{10.6}{\um} and core-wall thickness of \qty{186}{\nm}. They reported the onset of ultraviolet light generation with pump pulse energy as low as \qty{125}{\nano\J}. This small-core antiresonant fibre was manufactured by post-tapering a larger fibre. Only the constant cross-section taper waist was used. Very recently, Sabbah \etal{} \cite{Sabbah2023GreenUV} demonstrated the use of a \qty{6}{\um} core diameter antiresonant fibre, also manufactured through post-tapering, which enabled resonant dispersive wave emission in the deep ultraviolet with just $\sim\qty{20}{\nano\J}$ pulses at \qty{515}{\nm}.

\subsection{\label{sec:ssfbs}Plasma-driven soliton self-frequency blue-shift}
The larger guidance windows of antiresonant fibres enable much larger soliton self-frequency shifting. While both a Raman soliton self-frequency red-shift and plasma soliton self-frequency blue-shift were observed in bandgap fibres, the absolute frequency shifts were moderate, $\sim\qty{10}{\THz}$, due to the narrow guidance. In antiresonant fibres both of these processes be significantly enhanced.

The first to be demonstrated was the plasma\footnote{Note that ``plasma'' in hollow fibres is usually partially ionised, as opposed to a fully ionised plasma, with ionisation fractions up to the few percent level, but usually much lower.} soliton self-frequency blue-shift, by H\"{o}lzer \etal{} in 2011~\cite{holzer_femtosecond_2011}. This effect arises from the lower refractive index of plasma due to the strong negative polarisability of free electrons. At sufficiently high pump intensities, plasma formation occurs very rapidly around the temporal peak of a soliton, leading to a fast drop of the refractive index, and consequent phase modulation---the sign of which leads to a frequency up-shift~\cite{wood_measurement_1991,rae_detailed_1992}. In the experiment of H\"{o}lzer \etal{} a shift in excess of \qty{125}{\THz} was achieved when pumping a \qty{26}{\um} argon-filled fibre with \qty{65}{\fs}, \qty{800}{\nm} pulses with up to \qty{9}{\uJ} of energy. An example of experimental results illustrating this effect, by Sabbah \etal{}~\cite{sabbahGenerationCharacterizationFrequency2023a}, is shown in Fig.~\ref{fig:antiresonant}(b).

The role of plasma formation and soliton effects were clearly identified by numerical simulations. A co-published paper by Saleh \etal{}~\cite{saleh_theory_2011} (and subsequently Ref.~\cite{saleh_understanding_2011}) described a theoretical model, and through perturbation theory explained the soliton blue-shift in a way directly analogous, but opposite in sign, to the Raman soliton self-frequency red-shift familiar in solid-core optical fibres, along with further insight, such as temporal and spectral non-local soliton clustering, mediated by long-lived plasma. This phenomenon is a clear example of dynamics that can only occur when combining soliton dynamics with gas-based nonlinear optics.

The plasma soliton self-frequency blue-shift is important in many nonlinear optics experiments in hollow-core fibres. The vacuum ultraviolet resonant dispersive wave emission and supercontinuum formation demonstrated by Ermolov \etal{}~\cite{ermolov_supercontinuum_2015} were found to be strongly enhanced by the plasma blue-shift. In a different scenario in 2017, Tani \etal{} used blue-shifting self-compressed solitons to drive extreme ultraviolet high-harmonic generation \cite{taniContinuouslyWavelengthtunableHigh2017}. The extreme ultraviolet harmonics were frequency tuned by controlling the degree of soliton shift. The plasma induced after extreme soliton self-compression can also alter the dispersion landscape, such that an additional resonant dispersive wave can be phase-matched in the mid-infrared. This effect was predicted by Novoa \etal{} in 2015~\cite{novoa_photoionization-induced_2015}, and subsequently demonstrated by Köttig \etal{} in 2017 \cite{kottig2017novel}.

Soliton-plasma interaction has been studied in further detail by Huang \etal{} \cite{huangIonizationinducedAdiabaticSoliton2019,huangContinuouslyWavelengthtunableBlueshifting2019,huangHighlytunableVisibleUltrashort2019}. Of particular note was the demonstration of ionisation induced adiabatic soliton compression when starting with \qty{20}{\fs} pump pulses at \qty{800}{\nm}, with nearly all energy up-converted. These dynamics were predicted by Chang \etal{} in 2013 \cite{chang_combined_2013}.

Until recently a full time-frequency characterisation of the blue-shifting solitons had not been performed, with the soliton nature of the up-shifting pulse being inferred from numerical simulations and theory---both of which agree exceedingly well with experiment. The work of Tani \etal{} included some initial characterisation \cite{tani_wavelength-tunable_2017}, but did not clearly identify solitons. Very recently Sabbah \etal{}~\cite{sabbahGenerationCharacterizationFrequency2023a} used sum-frequency generation time-domain ptychography to fully characterise soliton self-compression and subsequent soliton self-frequency blue-shift for the first time. They observed self-compression of \qty{220}{\fs} pulses at \qty{1030}{\nm} to \qty{13}{\fs} in a \qty{2}{\m} long, \qty{29}{\um} core diameter, argon-filled fibre. A further increase in gas pressure led to blue-shifting sub-pulses being emitted. These shifted down to \qty{700}{\nm}, had a pulse duration of \qty{15}{\fs}, and propagated without chirp.

Beyond plasma soliton self-frequency blue-shift, the role of photoionisation and plasma formation can cause other effects in hollow-core fibre. In 2017 Köttig \etal{}~\cite{kottig2017phz}, discovered a plasma-induced soliton-fission process for pulses intense enough for ionisation to be important right from the start of the fibre, rather than after self-compression. The pulse can also undergo multiple stages of pulse compression and soliton fission \cite{SelimHabib:17}. In a separate theoretical work, Saleh \etal{} \cite{SalehMI2012} suggested a route to a new kind of modulational instability due to photoionisation.

In 2018, Koehler \etal{} \cite{koehlerLongLivedRefractiveIndexChanges2018,traboldSpatiotemporalMeasurementIonizationinduced2019}, found that the refractive-index changes caused by photoionisation can be very long-lived, up to tens of microseconds, and that they can lead to mechanical vibrations of the glass structures surrounding the core. As a result, a build-up of this effect can occur if subsequent pulses arrive within this timescale, corresponding to tens of kilohertz repetition rates. This effect was further investigated by Suresh \etal{}~\cite{sureshPumpProbeStudyPlasma2019}, who used counter-propagating solitons to probe the plasma density created at the temporal focus of a self-compressing higher-order pump soliton by observing changes in the emitted resonant dispersive wave wavelength. This build-up process can limit the scaling of nonlinear dynamics in antiresonant fibres to higher repetition rates for some parameter regimes. For example, Sabbah \etal{} observed the suppression of soliton self-frequency blue-shift at a repetition rate of \qty{200}{\kHz}~\cite{sabbahGenerationCharacterizationFrequency2023a}.

\subsection{\label{sec:ssfrs}Raman-driven soliton self-frequency red-shift}
Despite being very well established in solid-core fibres, and observed in bandgap fibres, the Raman-driven soliton self-frequency red-shift was not observed in antiresonant fibre until last year. Instead, the first interaction between the Raman response of molecular gases and solitons was observed in 2015 by Belli \etal{}~\cite{belli_vacuum-ultraviolet_2015}. In those experiments, a \qty{50}{\fs} pump pulse underwent Raman-enhanced soliton-effect self-compression, leading to vacuum ultraviolet dispersive wave emission. As the dispersive wave trailed behind the driving pulse, due to differences in group velocity, it was modulated by the Raman coherence impulsively excited in the gas by the driving pulse, leading to further broadening in the vacuum ultraviolet. This result was the first demonstration of vacuum ultraviolet guidance, the first vacuum ultraviolet supercontinuum, and the first demonstration of Raman-soliton interactions, in antiresonant fibre. Further investigation of the interaction between an ultrafast-pulse-induced Raman coherence and a probe pulse was discussed by Belli \etal{} in 2018~\cite{belliControlUltrafastPulses2018}.

While initial numerical simulations by Saleh \etal{} predicted Raman-soliton self-frequency red-shift in antiresonant fibres in 2015~\cite{saleh_tunable_2015}, predicting tuneable solitons between \qty{1300}{\nm} and \qty{1700}{\nm}, this was not experimentally demonstrated until the work of Chen \etal{} in 2022~\cite{chenEfficientSolitonSelffrequency2022a}. They used \qty{40}{\fs} pulses from a ytterbium laser system to pump a \qty{2}{\m} long, \qty{30}{\um} core diameter, hydrogen-filled antiresonant fibre and obtained Raman solitons tuneable between \qty{1080}{\nm} and \qty{1600}{\nm} with energy up to \qty{110}{\nano\J}. Subsequently, Tani \etal{} demonstrated tuneable pulses from the Raman soliton self-frequency shift across \qty{1100}{\nm} to \qty{1474}{\nm} at \qty{8}{\MHz} repetition rate in a \qty{1.6}{\m} long, \qty{40}{\um} core diameter, hydrogen-filled fibre pumped with \qty{220}{\nano\J}, \qty{40}{\fs} pulses~\cite{Tani2022}. This result is illustrated in Fig.~\ref{fig:antiresonant}(c). Furthermore, Tani \etal{} temporally characterised the frequency shifting pulses, finding them to be as short as \qty{22}{\fs}.

\section{\label{sec:hisol}Optical solitons in hollow capillary fibres}
In an effort to scale the peak power of soliton dynamics in antiresonant fibres, I came to the realization, in 2014, that we could in fact make use of hollow-capillary fibres without any microstructure at all. The key requirement for energy scaling is to increase the core size of the fibre. The maximum intensity that can be guided for nonlinear effects in a hollow-core fibre is limited by ionisation of the filling gas. While evacuated hollow capillary fibres can guide over \qty{e17}{\W\per\cm^2} \cite{Cros2004}, and Lekosiotis \etal{} have recently shown that much smaller core antiresonant fibres can guide at least \qty{3e15}{\W\per\cm^2} \cite{lekosiotis2023ontarget}, if you want to access soliton dynamics, there must be gas inside the fibre. Taking the lightest noble gas, with the highest ionisation potential, helium, we are limited to intensities of around $I_\mathrm{th}\sim\qty{3e14}{\W\per\cm^2}$ before excessive ionisation occurs. At this level, nonlinear dynamics are well-behaved, but at higher intensities, excessive ionisation of the filling gas consumes too much of the pulse energy and causes such huge phase modulation and spatial coupling, that the soliton dynamics break down. Therefore, in order to scale the peak power we must scale the core size via
\begin{equation}
A_\mathrm{eff}=1.5a^2 > \frac{P_0}{I_\mathrm{th}}.
\end{equation}

I realized that as the core size increases, antiresonant fibres become increasingly similar to simple hollow capillaries. To a good approximation, the dispersion is the same (unless one uses the resonances for dispersion enhancement), and so as we increase the core size we would be battling with low dispersion in both cases. And at larger core sizes, the attenuation in hollow capillaries drops rapidly and becomes tolerable. Therefore, I developed the scaling rules, Eq.~\ref{eqn:hcfloss} and Eq.~\ref{eqn:hcffiss}. From these, it is clear that for sufficiently large core size, it is always possible to make the soliton fission length shorter than the loss length, because the former scales as $a^2$ whereas the latter scales with $a^3$. The main problem is that the length can get very long. This is always the case in scaling any nonlinear optics experiment in gases, as described by Heyl \etal{}~\cite{heylScaleinvariantNonlinearOptics2016}. My solution was to use shorter pump pulses, which proportionally reduces the fission length, and to use the technique of stretching flexible capillaries to provide longer lengths than usual~\cite{nagyFlexibleHollowFiber2008}. As we will see, this approach was very successful, and I refer to it as HISOL (for high-energy solitons)---the acronym of the ERC Starting Grant project that funded this work.

It is worth considering why this was not realized before. For myself, it was because I was focused on lower energies at the few microjoule level. In that case, achieving sufficient nonlinearity dictates that a small core size is required, and hence antiresonant fibre must be used~\cite{travers_ultrafast_2011,russell_hollow-core_2014}. When using larger core sizes, as noted in Section~\ref{sec:presolhcf}, and Fig.~\ref{fig:fisslong}, for the usual parameters used in hollow-capillary post-compression experiments, soliton dynamics are inaccessible. Only by going through the development and analysis of optimising soliton dynamics in antiresonant fibre did I fully understand the scaling rules that enabled the conceptualisation of HISOL (Eq.~\ref{eqn:hcfloss} and Eq.~\ref{eqn:hcffiss}). While we have now shrunk HISOL systems down even to \qty{15}{\cm} (see below), within reach of rigid capillary systems, it is also important to note that when first devising this technique, the liberation of thought facilitated by being able to use arbitrarily long hollow capillaries---enabled by stretching flexible hollow capillaries---was a decisive factor in developing this idea. The technique that Nagy \etal{} developed not only enables very long fibre lengths (up to \qty{6}{\m} so far \cite{jeongDirectCompression170fs2018}), but also improves beam quality and transmission. With careful alignment and a good laser system we achieve performance at the theoretical limit, with up to 97\% total transmission in the fundamental mode, and my group uses our own stretching technique for all fibre lengths, even just a few cm, as the performance is so much improved compared to rigid capillaries.

There had been some suggestions to obtain soliton propagation in hollow capillary fibres before I developed the HISOL approach. In 2009 Husakou \etal{} suggested using dielectric-coated metal capillaries to enhance the guidance and allow the use of smaller core sizes in order to obtain soliton dynamics~\cite{husakouSolitoneffectPulseCompression2009}, before turning to antiresonant fibres~\cite{im_soliton_2010}. Another approach suggested was to pump in higher order modes, which have larger dispersion~\cite{Lopez-Zubieta2018}. In a similar vein, one can move to much longer wavelengths, into the mid-infrared, where dispersion also increases~\cite{Voronin2014,Zhao2017}. However, these were all numerical results, and lacked the simplicity of using readily available fibres, at accessible pump wavelengths, as well as the convenience and high beam quality obtained by making use of the fundamental mode. The closest experimental result was perhaps a cascaded compression setup~\cite{schenkelGeneration8fsPulses2003a}, where \qty{10}{\fs} pulses produced in a first fibre were used to pump a second fibre to generate a supercontinuum. However, the dispersion landscape, set by the core size and gas pressure combination used, was not suitable for the observation of soliton dynamics.

\begin{figure}[tb!]
    \centering
    \includegraphics{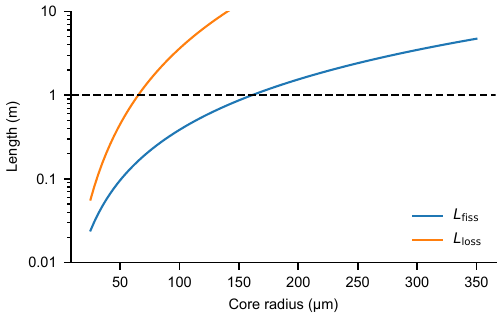}
    \caption{\label{fig:fissshort}\textbf{Soliton fission length in hollow capillary fibres with short pump pulses.} (Compare to Fig.~\ref{fig:fisslong}.) The soliton fission length $L_\mathrm{fiss}$ and loss length $L_\mathrm{loss}$ as a function of core radius for a hollow capillary filled with helium such that $\lambda_\mathrm{zd}=\qty{600}{\nm}$, pumped with \qty{8}{\fs} pulses at \qty{800}{\nm}. For each core size, the gas pressure is scaled to set $\lambda_\mathrm{zd}=\qty{600}{\nm}$, and the pump peak power is set to the maximum allowed intensity, following the limits determined in \cite{Travers2019}. For all core sizes $L_\mathrm{fiss} < L_\mathrm{loss}$, so soliton dynamics are accessible, furthermore, $L_\mathrm{fiss}$ is short enough to be practical even for relatively large core sizes, enabling very high peak power to be used.}
\end{figure}

In Fig.~\ref{fig:fissshort} I plot the soliton fission length $L_\mathrm{fiss}$ and loss length $L_\mathrm{loss}$ as a function of core radius for a hollow capillary filled with helium such that $\lambda_\mathrm{zd}=\qty{600}{\nm}$, pumped with \qty{8}{\fs} pulses at \qty{800}{\nm}. When using shorter pump pulses it is clear that $L_\mathrm{fiss} < L_\mathrm{loss}$ for all core sizes, so that soliton dynamics are accessible. From the figure it is also clear that $L_\mathrm{fiss}$ is short enough to be practical even for relatively large core sizes, enabling very high peak power to be used. (Compare this to Fig.~\ref{fig:fisslong}.) Furthermore, we can still obtain similar dispersion-landscape tuning as with antiresonant fibre, as illustrated in Fig.~\ref{fig:disptune}(b). It is straightforward to tune $\lambda_\mathrm{zd}$ across the deep ultraviolet, visible and near-infrared region. The values of dispersion in a large-core hollow capillary are of course much lower than a small-core antiresonant fibre; as an example for a \qty{250}{\um} core diameter hollow capillary, filled with \qty{1}{\bar} helium, the group velocity dispersion at \qty{800}{\nm} is just \qty{-7.6}{\fs^2\per\m}, but when using short, \qty{8}{\fs}, pump pulses the dispersion length is a reasonable \qty{2.7}{\m}. The nonlinear coefficient in this case is just $\sim\qty{1.2e-10}{\per\W\per\m}$ at \qty{800}{\nm}, but this low value is compensated by the fact that we can greatly scale the peak pump power.

\begin{figure*}[t!]
    \centering
    \includegraphics{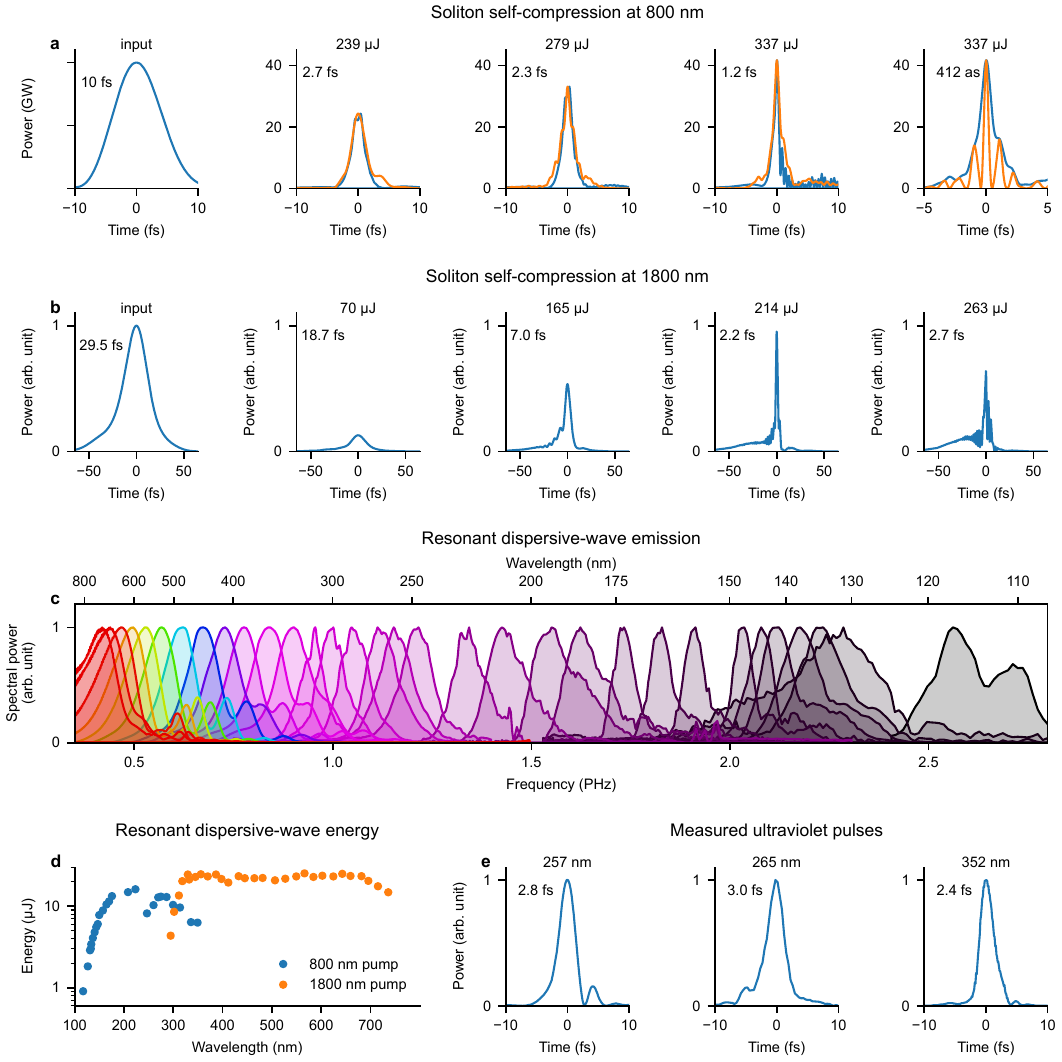}
    \caption{\label{fig:HISOL} \textbf{Soliton results in hollow capillary fibres.} (a) Soliton self-compression of \qty{10}{\fs}, \qty{800}{\nm} pulses to a sub-cycle envelope duration of \qty{1.2}{\fs}, with \qty{43}{\GW} peak power in a \qty{3}{\m} long, \qty{250}{\um} core diameter hollow capillary, filled with \qty{0.4}{\bar} helium (data from \cite{Travers2019}). (b) Soliton self-compression of \qty{30}{\fs}, \qty{1800}{\nm} pulses to a sub-cycle envelope duration of \qty{2}{\fs}, with \qty{27}{\GW} peak power in a \qty{2.5}{\m} long, \qty{450}{\um} core diameter hollow capillary filled with \qty{0.83}{\bar} argon (data from \cite{brahmsInfraredAttosecondField2020}). (c) Tuneable resonant dispersive wave emission across the vacuum and deep ultraviolet (data from \cite{Travers2019} and \cite{brahmsInfraredAttosecondField2020}). (d) Ultraviolet pulse energy corresponding to (c). (e) Measured ultraviolet pulse durations at several wavelengths when using a \qty{150}{\um} core diameter, \qty{60}{\cm} long, neon filled fibre, driven by \qty{10}{\fs}, \qty{800}{\nm} pulses (data from \cite{reduzziDirectTemporalCharacterization2023}).}
\end{figure*} 

Based on this insight, Teodora Grigorova, Federico Belli and I experimentally demonstrated the HISOL concept for the first time on 10$^\mathrm{th}$ November 2017~\cite{Grigorova:18,Travers:18}. We made use of a compressed Ti:sapphire laser system, producing $\sim\qty{10}{\fs}$ pulses, to pump a \qty{3}{\m} long, helium-filled, stretched hollow capillary with \qty{250}{\um} core diameter. Over the following 18 months, Christian Brahms joined the team, and we refined the experimental setup, diagnostics, and analysis, and published our findings in detail~\cite{Travers2019}. The core results are shown in Fig.~\ref{fig:HISOL}. We explained the concept for peak-power scaling of optical solitons, and experimentally demonstrated two key effects for the first time in hollow capillaries: we observed soliton self-compression to sub-cycle (\qty{1.2}{\fs}) pulses with \qty{43}{\GW} peak power (Fig.~\ref{fig:HISOL}a), and resonant dispersive wave emission between \qty{110}{\nm} and $>\qty{350}{\nm}$ with energies of 1 to \qty{16}{\uJ}, Fig.~\ref{fig:HISOL}(c,d). Measurement of the \qty{1.2}{\fs} self-compressed pulse also represents the shortest optical pulse ever measured by a frequency-resolved optical gating technique, and was actually the hardest part of the experiment.

These results represent a remarkable energy and peak-power scaling of soliton dynamics in hollow fibre. The sub-femtosecond pulse self-compression provides a very simple route to high peak power optical attosecond pulse generation~\cite{Hassan2016}, which enables new attosecond interactions with matter with non-ionising photon energies. Resonant dispersive wave emission in hollow capillaries also offers the highest conversion efficiencies---by multiple orders of magnitude---of infrared pulses to the vacuum ultraviolet spectral region yet demonstrated, leading to the highest energy tunable source of few-femtosecond pulses across this spectral region. The vacuum and deep ultraviolet pulse generation offers a single source covering 3--12~eV with multiple-\unit{\uJ} energies and few-femtosecond duration, also in an extremely simple setup, opening up new investigations in fundamental science---for example as a versatile pump source in a multitude of experiments in ultrafast science~\cite{kotsinaUltrafastMolecularSpectroscopy2019,teles-ferreira_ultrafast_2022,maiuri_ultrafast_2020,chergui_ultrafast_2019,Wanie2021, kotsinaSpectroscopicApplicationFewfemtosecond2022,calegari2023open}---and potentially new industrial applications.

The HISOL technique also enables, for the first time, fibre soliton dynamics to be harnessed at peak power and energy levels comparable to the state-of-the-art from ultrafast laser technology. The use of hollow capillaries also liberates us from avoiding the resonances in antiresonant fibres, providing uninterrupted guidance and smooth dispersion. Note that these results are based on almost pure fundamental mode propagation, so the resulting self-compressed or ultraviolet pulses have excellent beam quality~\cite{Travers2019}.

Over the subsequent years we widely explored this platform. In 2019, Brahms \etal{}~\cite{brahmsHighenergyUltravioletDispersivewave2019} demonstrated that they could significantly shrink this system through a combination of even shorter pump pulses and a smaller core radius. Using \qty{6.3}{\fs} pump pulses and \qty{100}{\um} core diameter, a hollow capillary just \qty{15}{\cm} long was sufficient to obtain resonant dispersive waves tuneable from \qty{218}{\nm} to \qty{375}{\nm}, with more than \qty{4}{\uJ} output energy across the whole band. When using a \qty{34}{\cm} long, \qty{150}{\um} core diameter hollow capillary, this energy scaled to $\sim\qty{10}{\uJ}$ across the same band.

In 2020, Brahms \etal{} demonstrated that all the key features we had seen with an \qty{800}{\nm} pump wavelength were reproduced when pumping further into the infrared at \qty{1800}{\nm}~\cite{brahmsInfraredAttosecondField2020}. This included soliton self-compression of \qty{30}{\fs} pulses to a sub-cycle envelope duration of \qty{2}{\fs}, with \qty{27}{\GW} peak power (see Fig.~\ref{fig:HISOL}(b)), and resonant dispersive wave emission from \qty{300}{\nm} to \qty{740}{\nm} with energy up to \qty{25}{\uJ} and efficiency up to 12\% (Fig.~\ref{fig:HISOL}(c,d)), both in a \qty{2.5}{\m} long, \qty{450}{\um} core diameter hollow capillary filled with argon. The enhanced dispersion at longer wavelengths actually makes soliton dynamics more accessible, and so there was no need to post-compress the \qty{30}{\fs}, \qty{1800}{\nm} pulses from our optical parametric amplifier in order to drive the soliton dynamics. Achieving such short pulse durations (\qty{2}{\fs}) in the shortwave infrared region---which we called infrared attosecond pulses, as they corresponded to a sub-femtosecond electric field transient---will simplify driving high-harmonic generation of isolated attosecond pulses in the soft X-ray region. In addition, one of the most important results in that paper was the full characterisation of the spectral-temporal dynamics of both the self-compression process, and the resonant dispersive wave emission, for the first time~\cite{brahmsInfraredAttosecondField2020}. When pumping a more compact second stage---\qty{38}{\cm} long, \qty{200}{\um} core diameter---with shorter pulses (\qty{16}{\fs}) the resonant dispersive wave tuning was extended down to \qty{210}{\nm}. Further consideration of the wavelength scaling of the HISOL concept was reported in~\cite{cregoToolsNumericalModelling2023}, including scaling rules and a sophisticated numerical propagation model.

The use of pressure gradients for soliton dynamics, first pioneered by Mak \etal{} in antiresonant fibre \cite{mak_tunable_2013}, was also demonstrated in hollow capillary fibres~\cite{brahmsResonantDispersiveWave2020}. The use of pressure gradients is critical for the delivery of ultrashort ultraviolet pulses to vacuum, due to the high dispersion of any possible window material. We further discovered, from numerical simulations, that when using a decreasing pressure gradient the few-femtosecond ultraviolet pulses are delivered to the end of the fibre (and hence to the experiment) at close to their transform-limited duration. A simple $3/2$ end-pressure scaling rule was found that enables dynamics to be transferred between a static pressure and gradient pressure configuration, preserving the resonant dispersive wave emission wavelength. While this scaling rule is obvious when considering total nonlinearity, it was surprising that it worked even for resonant dispersive wave emission, which is a highly dynamic and transient phenomenon that strongly depends on dispersion. Additional numerical studies of the optimal pulse self-compression conditions have also been reported in detail~\cite{galanOptimizationPulseSelfcompression2022,galanScalableSubcyclePulse2023}.

Delivery of short pulses to vacuum through a decreasing pressure gradient was recently confirmed experimentally by Reduzzi \etal{} \cite{reduzziDirectTemporalCharacterization2023}, who used an all-in-vacuum self-diffraction frequency-resolved optical gating setup, to directly measure the generated ultraviolet pulses without additional dispersion. Resonant dispersive waves tuneable between \qty{250}{\nm} and \qty{350}{\nm} were generated in a \qty{150}{\um} core diameter, \qty{60}{\cm} long, neon filled fibre, driven by \qty{10}{\fs}, \qty{800}{\nm} pulses. Fig.~\ref{fig:HISOL}(e) shows the measured temporal profiles for three example pulses, with pulse durations $<\qty{3}{\fs}$ measured across the range. Zhang \etal{} \cite{zhangMeasurementsMicrojoulelevelFewfemtosecond2022a}, instead directly measured resonant dispersive waves around \qty{400}{\nm} in air, with the help of dispersive mirrors, demonstrating \unit{\uJ}-level pulses with $\sim\qty{9.5}{\fs}$ duration.

In 2021, Brahms and Travers~\cite{brahmsTimingEnergyStability2021}, numerically studied the timing and energy stability of resonant dispersive wave emission in hollow capillary fibres in the presence of noise. We found that for low pump energy, fluctuations in the pump energy are strongly amplified. However, when the generation process is saturated, the energy of the ultraviolet pulse can be significantly less noisy than that of the pump pulse. Furthermore, we found that the arrival-time jitter of the generated pulses remains well below one femtosecond even for a conservative estimate of the pump pulse energy noise, and that photoionisation and plasma dynamics can aid the stability of the ultraviolet pulses.

Lekosiotis \etal{} experimentally and numerically demonstrated that all the dynamics described above can also be obtained with circular polarisation~\cite{lekosiotisUltrafastCircularlyPolarized2021}, simply by converting the pump pulse to circular polarisation. This is a result of the negligible birefringence in stretched hollow capillary fibres. Specifically, Lekosiotis \etal{} demonstrated circularly polarized pulses tunable from \qty{160}{\nm} to \qty{380}{\nm} with up to \qty{13}{\uJ} of pulse energy. We also numerically verified that a simple $3/2$ scaling of pulse energy between linearly and circularly polarized pumping closely reproduces the soliton and dispersive wave dynamics---as with the case of pressure gradients, this rule seems simple in retrospect (and well established in the case of nonlinear spectral broadening), but was unexpected given the complex nature of the soliton dynamics.

In 2022, Brahms and Travers demonstrated soliton self-compression and ultraviolet resonant dispersive wave emission when pumping in a higher-order mode~\cite{brahmsSolitonSelfcompressionResonant2022}. Specifically, we excited the double-lobe LP$_{11}$ mode of an argon-filled hollow capillary and observed the generation of ultraviolet resonant dispersive waves that were frequency-shifted and more narrowband compared to fundamental mode pumping, due to the stronger modal dispersion. It was also observed that the supercontinuum between the resonant dispersive wave and pump pulse was greatly suppressed due to a larger phase-mismatch resulting from the larger dispersion. Furthermore, due to the lower field strength in some higher-order modes for a given peak power, it was found numerically that photoionisation and plasma effects could also be suppressed, potentially allowing for higher pulse energy to be used in the self-compression process. An additional numerical paper considered self-compression of pulses propagating in higher-order modes in pressure gradients~\cite{wanEffectDecreasingPressure2021}.

Several applications of HISOL-based sources have already been demonstrated. In 2022, Kotsina \etal{} applied HISOL-based ultraviolet sources to photoelectron spectroscopy~\cite{kotsinaSpectroscopicApplicationFewfemtosecond2022}. Based on photoelectron cross-correlation measurements they estimated a deep ultraviolet pulse duration of \qty{6}{\fs} in their experiments. Also in 2022, Fazio \etal{} \cite{fazioSolitonSelfcompressionBased2022}, used self-compressed pulses from a hollow capillary to drive high-harmonic generation. The availability of few-femtosecond tuneable vacuum and deep ultraviolet pulses, with high energy, opens the door to many experiments that were previously very challenging, such as ultraviolet driven attochemistry~\cite{calegari2023open}.

Recently, efforts have been made to switch to ytterbium pump laser technology in order to increase the repetition rate, and to simplify the laser technology, to enable transfer of these sources outside of research laboratories. First, Brahms and Travers used a two-stage system starting with a compact and industrialised ytterbium laser producing \qty{220}{\fs} pulses~\cite{brahmsEfficientCompactSource2023}. In the first stage, a hollow capillary post-compression setup reduced the pulse duration to \qty{13}{\fs}. In the second stage, soliton self-compression and resonant dispersive wave emission were optimised. The system was designed to be compact and occupied a total footprint of only $\qty{120}{\cm}\times\qty{75}{\cm}$, including the pump laser. Ultraviolet pulses between \qty{208}{\nm} and \qty{363}{\nm} were generated at \qty{50}{\kHz}, with average powers up to \qty{0.36}{\W} and a total pump-laser to ultraviolet efficiency up to 3.6\%. Brahms and Travers then extended this concept to a three-stage system~\cite{Brahms2023b}, compressing a higher energy ytterbium laser with \qty{330}{\fs}, \qty{950}{\uJ} pulses, first to \qty{28}{\fs} in a \qty{1.75}{\m} long \qty{530}{\um} core diameter argon-filled fibre, then to \qty{8.3}{\fs} in a \qty{1.4}{\m} long \qty{530}{\um} core diameter argon-filled fibre, and then used those pulses to generate tuneable vacuum ultraviolet emission down to \qty{140}{\nm} in a \qty{0.5}{\m} long, \qty{150}{\um} core diameter, helium-filled fibre, at \qty{4}{\kHz} repetition rate.

In a different configuration, Silletti \etal{} demonstrated the use of a dispersion-controlled multi-pass cell as the first stage compressor, further reducing the system footprint~\cite{Silletti2023}. In the multi-pass cell compressor they reduced the \qty{150}{\fs}, \qty{1030}{\nm} pump pulses down to $\sim\qty{17}{\fs}$ pulses with \qty{230}{\uJ} of energy. These pulses were then used to drive resonant dispersive wave emission in a hollow capillary filled with argon, achieving tuneability between \qty{260}{\nm} and \qty{530}{\nm} at a \qty{10}{\kHz} repetition-rate. 

So far we have discussed soliton self-compression and resonant dispersive wave emission. However, the use of soliton and related dynamics in hollow capillaries opens the door to a broader range of phenomena. In a wide-ranging experimental and numerical paper, Grigorova \etal{} systematically surveyed multiple dispersion regimes for pulse propagation in hollow capillary fibres~\cite{grigorovaDispersiontuningNonlinearOptical2023}. As established already, when the pump pulse propagates in the anomalous dispersion regime (achieved at relatively low pressure) they observed soliton dynamics accompanied by soliton-plasma effects, such as self-compression, resonant dispersive wave emission in the fundamental as well as in higher-order modes, soliton blue-shifting and ionisation-induced pulse splitting. By increasing the pressure, propagation of the pump pulse in the vicinity of the zero-dispersion wavelength was investigated and observed to result in pulse splitting and subsequent cross-phase modulation, leading to the generation of an additional frequency-shifted band and a three-octave broad supercontinuum. At yet higher pressure, when the pump pulse is in the normal dispersion regime, they observed the generation of a broad and flat supercontinuum. In this regime, the experimental results are less well described by simulations that consider only the propagation dynamics inside the fibre. Free-space simulations of the beam propagation in the bulk gas, before the capillary entrance, suggest that this discrepancy is caused by self-focusing and ionisation altering the pulse spatial and temporal shape, affecting both the coupling efficiency and the subsequent propagation inside the capillary.

In summary, there is a rich range of nonlinear dynamics to be explored when parameters are chosen such that dispersion in hollow capillary fibres is not negligible. While the soliton results already shown are exciting and useful, there is much more to be done. In particular, further repetition rate and average power scaling with more advanced ytterbium laser technology seem inevitable, especially with an eye to increasing the average power in the deep and vacuum ultraviolet. I expect more work to make these systems more compact, and also towards pumping these dynamics both further into the infrared, and also at shower wavelengths. It is clear that ultrafast soliton dynamics in hollow capillaries have a bright future.

\section{Summary and future perspectives}
Solitons in hollow-core optical fibres have provided a platform for a new class of light sources, including extreme pulse compression, tuneable and few-cycle ultraviolet to visible pulse generation through resonant dispersive wave emission, and tuneable visible and infrared light sources via soliton shifting. These sources often produce light with unique properties that cannot be obtained by other means, with consequent wide-ranging applications in both fundamental science and industry. A big question is: what is next? Predicting the future is a futile task, and I will not attempt it. Instead, here are some directions that remain unexplored.

There are some fundamental questions that remain unanswered. How short in wavelength can dispersive wave generation really extend below \qty{100}{\nm}? What is the shortest duration that a self-compressed pulse can be? Are there any limits to energy scaling? The recent demonstration by Sabbah \etal{} in a tiny core---\qty{6}{\um} diameter---fibre required just \qty{20}{\nano\J} pump pulses~\cite{Sabbah2023GreenUV}. Can this be pushed even lower? At the other extreme are plans to finally realise soliton dynamics at the terawatt peak power level, as predicted in our original HISOL paper~\cite{Travers2019}. If realised, we would obtain a terawatt peak power optical attosecond pulse source, capable of driving relativistic nonlinear optics at sub-cycle timescales for the first time. The associated vacuum ultraviolet pulses could reach peak powers of $>\qty{100}{\GW}$ and energies $>\qty{100}{\uJ}$, sufficient to drive few-femtosecond strong-field physics at vacuum ultraviolet photon energies.

There are also areas that have received somewhat less attention than others. For example, the use of pump wavelengths even further into the infrared is not well explored---can we drive soliton dynamics beyond \qty{10}{\um} wavelength, or even further in the THz region? Furthermore, the full dimensionality of light has not yet been fully harnessed. Much remains to be explored in terms of polarisation evolution and control---beyond linear and circular polarisation---and in terms of multimode dynamics. Nearly everything I have discussed in this paper is predominantly fundamental mode, or a deliberately selected few higher order modes. But there is growing interest in fully multimode soliton-like dynamics, such as the recently investigated multidimensional solitary states \cite{safaeiHighenergyMultidimensionalSolitary2020}, and spatial cage solitons~\cite{Mei:22}---these are completely different to the fundamental-mode soliton dynamics I have discussed in this article.

Finally, it is the case that applications are only just emerging. The sub-femtosecond visible-infrared pulses, and the tuneable few-femtosecond ultraviolet sources, that I have described, are still quite young, and I am aware of many efforts to integrate these sources into fundamental science experiments in institutes around the world. Beyond the laboratory, it is clear that if made practical, such sources can also be useful for both healthcare and industrial applications, where the unique properties of deep and vacuum ultraviolet light, in particular, enable new light-matter interactions and hence new applications.

We have come a very long way since Russell's first observations of solitons in the Union Canal, and even since Hasegawa and Tappert's study of solitons in optical fibres. Despite this, the science of solitons is still a very active and exciting field of research, and soliton dynamics in hollow-core fibres are at the frontier of this research effort.

\section*{Acknowledgements}
I thank my whole research group (\url{https://lupo-lab.com}), who work daily to advance the field of optical soliton effects in hollow-core fibres. I thank Christian Brahms, Leah Murphy, Mohammed Sabbah, Francesco Tani, Martin Gebhardt, Mohammed F. Saleh, Jonathan Knight, and Teodora Grigorova, for carefully reading and providing feedback on this manuscript, and Francesco Tani for supplying the data for Fig.~\ref{fig:antiresonant}(c). This work was part-funded by the European Research Council (ERC) under the European Union Horizon 2020 Research and Innovation program: Consolidator Grant agreement XSOL, No.~101001534, and by the Institution  of  Engineering  and  Technology (IET) through the IET A F Harvey Engineering Research Prize.

\end{document}